\documentclass[12pt]{iopart}
\usepackage{latexsym}
\usepackage{amscd}
\begin{document}

 \def\lambdabar{\protect\@lambdabar}
\def\@lambdabar{%
\relax
\bgroup
\def\@tempa{\hbox{\raise.73\ht0
\hbox to0pt{\kern.25\wd0\vrule width.5\wd0
height.1pt depth.1pt\hss}\box0}}%
\mathchoice{\setbox0\hbox{$\displaystyle\lambda$}\@tempa}%
{\setbox0\hbox{$\textstyle\lambda$}\@tempa}%
{\setbox0\hbox{$\scriptstyle\lambda$}\@tempa}%
{\setbox0\hbox{$\scriptscriptstyle\lambda$}\@tempa}%
\egroup
}

\def\bbox#1{%
\relax\ifmmode
\mathchoice
{{\hbox{\boldmath$\displaystyle#1$}}}%
{{\hbox{\boldmath$\textstyle#1$}}}%
{{\hbox{\boldmath$\scriptstyle#1$}}}%
{{\hbox{\boldmath$\scriptscriptstyle#1$}}}%
\else
\mbox{#1}%
\fi
}
\def\msf{\hbox{{\sf M}}}
\def\psf{\hbox{{\sf P}}}
\def\Nsf{\hbox{{\sf N}}}
\def\Tsf{\hbox{{\sf T}}}
\def\Asf{\hbox{{\sf A}}}
\def\Bsf{\hbox{{\sf B}}}
\def\Lsf{\hbox{{\sf L}}}
\def\Ssf{\hbox{{\sf S}}}
\def\Mtens{\bi{M}}
\def\msfsim{\bbox{{\sf M}}_{\scriptstyle\rm(sym)}}
\newcommand{\mcsim}{ {\sf M}_{ {\scriptstyle \rm {(sym)} } i_1\dots i_n}}
\newcommand{\mcs}{ {\sf M}_{ {\scriptstyle \rm {(sym)} } i_1i_2i_3}}

\newcommand{\beqan}{\begin{eqnarray*}}
\newcommand{\eeqan}{\end{eqnarray*}}
\newcommand{\beqa}{\begin{eqnarray}}
\newcommand{\eeqa}{\end{eqnarray}}

 \newcommand{\suml}{\sum\limits}
\newcommand{\intl}{\int\limits}
\newcommand{\rvec}{\bbox{r}}
\newcommand{\xivec}{\bbox{\xi}}
\newcommand{\Avec}{\bbox{A}}
\newcommand{\Rvec}{\bbox{R}}
\newcommand{\Evec}{\bbox{E}}
\newcommand{\Bvec}{\bbox{B}}
\newcommand{\Svec}{\bbox{S}}
\newcommand{\avec}{\bbox{a}}
\newcommand{\nablav}{\bbox{\nabla}}
\newcommand{\nuvec}{\bbox{\nu}}
\newcommand{\bvec}{\bbox{\beta}}
\newcommand{\vvec}{\bbox{v}}
\newcommand{\jvec}{\bbox{j}}
\newcommand{\nvec}{\bbox{n}}
\newcommand{\pvec}{\bbox{p}}
\newcommand{\mvec}{\bbox{m}}
\newcommand{\evec}{\bbox{e}}
\newcommand{\eps}{\varepsilon}
\newcommand{\la}{\lambda}
\newcommand{\rad}{\mbox{\footnotesize rad}}
\newcommand{\scr}{\scriptstyle}
\newcommand{\latens}{\bbox{\Lambda}}
\newcommand{\pitens}{\bbox{\Pi}}
\newcommand{\cm}{{\cal M}}
\newcommand{\cp}{{\cal P}}
\newcommand{\beq}{\begin{equation}}
\newcommand{\eeq}{\end{equation}}
\newcommand{\ptens}{\bbox{\sf{P}}}
\newcommand{\Ptens}{\bbox{P}}
\newcommand{\Ttens}{\bbox{\sf{T}}}
\newcommand{\Ntens}{\bbox{\sf{N}}}
\newcommand{\Ncal}{\bbox{{\cal N}}}
\newcommand{\Atens}{\bbox{\sf{A}}}
\newcommand{\Btens}{\bbox{\sf{B}}}
\renewcommand{\d}{\partial}
\def\rmi{{\rm i}}
\def\rme{\hbox{\rm e}}
\def\rmd{\hbox{\rm d}}

\title{ Some Aspects of the Electromagnetic Multipole Expansions}
\author{Irina Dumitriu 
\footnote{Now at AG Moderne Optik, Institut f\"{u}r Physik, Humboldt-Universit\"{a}t zu Berlin, Hausvogteiplatz 5-7, D-10 117 Berlin, Germany, 
} 
and C. Vrejoiu \footnote{E-mail :  Irina.Dumitriu@physik.hu-berlin.de,
 cvrejoiu@yahoo.com }}
\address{Facultatea de Fizic\v{a}, Universitatea din Bucure\c{s}ti, 077 125, 
Bucure\c{s}ti-M\v{a}gurele, Rom\^{a}nia  }
\begin{abstract}
Various procedures for expressing the multipolar expansion of the electromagnetic 
field are considered with application to the calculation of the radiated power. 
Some results from the literature are discussed and  perspective  of developing 
the subject   is pointed out.
\end{abstract}

\section{Introduction}

 The multipole expansion of the electromagnetic field is an  usefull tool in physics and can be found in any book on electrodynamics or on the theory of atomic and nuclear transitions. In most of the textbooks (see, e.g., \cite{jack,La}) a full and systematic treatment is given in spherical coordinates, while for Cartesian coordinates the problem is presented fully only for the static case; the dynamic case is given only for the lowest order multipoles. A general procedure for the reduction of the multipole tensors represented by Cartesian coordinate components to fully symmetric traceless ones in the static  case is given in \cite{sc,Vr} and it  is generalized in \cite{vr} to the dynamic case.
This method is applied in \cite{VN,Di} to the radiation field.
\par In the present paper  different procedures are applied for calculating the
total power radiated by a confined system of charges. A first one represents the traditional
method of expressing the field expansions by the multipole Cartesian tensors, applying
finally the reduction of these tensors. The second one uses the reduction technique
done in \cite{vr} but more systematically considered here.
The results of \cite{VN,Di} are analysed and compared to those of other methods from the 
literature \cite{bel,ra}. The advantages of the Cartesian coordinates are emphasized 
firstly by the simplicity of the formalism : only algebraic manipulations and 
combinatorics are implied, no special functions being required.  Secondly, the procedure initiated in \cite{vr} and used here 
leads to a nontrivial grouping of different multipolar contributions standing out the 
toroidal multipole contributions \cite{dub1,dub2,dub3}
\par In section 2 some basic formulae for multipole expansions are presented. 
Section  3 deals with the radiation field as well as with the expression and 
expansion of the total radiated power. In Section 4 the procedure of reduction 
of a tensor to a symmetric traceless one is given. The total radiated power is 
then treated in Section 5 using these reduced moments and a comparison with  literature is made. 
Section 6 presents, in a more systematic and concise way, the results from 
\cite{VN,Di}. Then the total radiated power is expressed in Section 7  
by the transformed 
moments. The conclusions are given in Section 8. In Appendix the proof of 
some formulas used throughout this paper is given, as well as the reduction scheme 
and a justification of the reduction procedure using the charge and current expansions.

\section{Basic Formulae for the Multipole Expansions}
Let us consider charge $\rho(\rvec,t)$ and current $ \jvec(\rvec,t)$  distributions
having supports included in a finite  domain  ${\cal D}$. Choosing the origin $O$ of
the Cartesian coordinates in ${\cal D}$, and using the notation
 $\evec_i$ for the orthogonal unit vectors along the axes, the retarded scalar
and vector potentials at a point outside ${\cal D}$,  $\rvec=x_i\evec_i$,  are
\begin{equation}\label{potret}
\fl\Phi(\rvec,t)=\frac{1}{4\pi\eps_0}\int \frac{\rho(\xivec,t-R/c)}{R}\rmd^3\xi,\;\;
\Avec(\rvec,t)=\frac{\mu_0}{4\pi }\int \frac{\jvec(\xivec,t-R/c)}{R}\rmd^3\xi
\end{equation}
where $\Rvec=\rvec-\xivec$. The Taylor series expansion of the function $f(R)$ is
\beqan
&~&\fl f(R)=\suml^\infty_{n=0}\frac{(-1)^n}{n!}\xi_{i_1}\dots \xi_{i_n}\d_{i_1
\dots i_n}f(r)=\suml^\infty_{n=0}\frac{(-1)^n}{n!}\xivec^n || \nablav^nf(r)
\eeqan
where
 $$\d_{i_1\dots i_n}=\frac{\d}{\d x_{i_1}}\dots \frac{\d}{\d x_{i_n}}$$
and $\bbox{a}^n$ is the $n$-fold tensorial product
$\bbox{a}\otimes \dots\otimes\bbox{a}:\;\;\left(\bbox{a}\otimes \dots\otimes\bbox{a}\right)_
{i_1\dots i_n}=a_{i_1}\dots a_{i_n}$.
Denoting by ${\Ttens}^n$ a {\it n}th order tensor, ${\Atens}^{(n)}||{\Btens}^{(m)}$ is a $|{\it n-m}|$th order tensor with the
components:
\beqan
\fl&~& ({\Atens}^{(n)}||{\Btens}^{(m)})_{i_1 \cdots i_{|n-m|}}=
\left\{\begin{array}{ll}
A_{i_1 \cdots i_{n-m}j_1 \cdots j_m}B_{j_1 \cdots j_m} & \;\rm{,}\; n>m\\
A_{j_1 \cdots j_n}B_{j_1 \cdots j_n} & \;\rm{,}\; n=m\\
A_{j_1 \cdots j_n}B_{j_1 \cdots j_n i_1 \cdots i_{m-n}} & \;\rm{,}\; n<m
\end{array} \right.
\eeqan
By applying the formula for the Taylor series expansion to the scalar
potential we get:
\begin{equation}\label{dfi}
\fl \Phi(\rvec,t)=\frac{1}{4\pi\eps_0}\suml^\infty_{n=0}\frac{(-1)^n}{n!}\nablav^n
|| \left[\frac{\ptens^{(n)}(t-r/c)}{r}\right],\;\;\ptens^{(n)}(t)=
\intl_{{\cal D}}\rvec^n\rho(\rvec,t)\rmd^3x,
\end{equation}
 $\ptens^{(n)}$ being the $n$-th order electric multipole  tensor.
\par For the vector potential we obtain the expression:
\begin{eqnarray}\label{expA1}
\fl{\Avec}(\rvec,t) &=& \frac{\mu_0}{4\pi}\suml^\infty_{n=0}
\frac{(-1)^n}{n!} \nablav^n ||
\left[\frac{\mu^{(n+1)}(t-r/c)}{r}\right]\nonumber\\
\fl&=& \frac{\mu_0}{4\pi}\suml^\infty_{n=0}
\frac{(-1)^n}{n!} \evec_i \d_{i_{1}} \cdots
\d_{i_{n}}\left[\frac{\mu_{{i_1} \cdots {i_n}i}(t_0)}{r}\right] \;\rm{,}\;\rm{t_0=t-r/c}
\end{eqnarray}
In the previous equation the magnetic multipole tensor was introduced by its
Cartesian components:
$$
\mu_{{i_1} \cdots {i_n}i} =  \int\limits_{{\cal
      D}} x_{i_1} \ldots x_{i_n} j_{i}(\rvec,t)d^{3}x
$$

\section{The Radiation Field}

The formula for the power radiated by a charged system described by the
charge $\rho$ and current $\jvec$ densities with supports included in a
finite domain ${\cal D}$ is well known \cite{La}:
\begin{equation}\label{angd}
{\mathcal J} (\nuvec) = \frac{\rmd{\it P}}{\rmd\Omega}(\nuvec,t)=
\frac{r^2}{\mu_{0}c}\left[\nuvec  \times \frac{\partial}{\partial t}\Avec_{rad}(\rvec,t)\right]^{2}
\end{equation}
Here, $\nuvec = \rvec/r$ and $\rmd{\it P}/\rmd\Omega$ is related to the
flow of the energy detected in the observation point $\rvec$ at large
distance $r$ compared with the dimensions of the given charged system.
The vector $\Avec_{rad}$ is obtained from the retarded potential \eref{potret}  by
retaining only the dominant terms at large distances.
\par In the following the expansion of $\Avec_{rad}$ is derived \cite{Vr,VN}. Starting
  from:
\beqan
\Avec(\rvec,t) &\approx&
\frac{\mu_0}{4\pi}\frac{1}{r}\suml^\infty_{n=0}\frac{(-1)^n}{n!}
\nablav ^n || \mu^{(n+1)}(t-r/c) \nonumber\\
&\approx& \frac{\mu_0}{4\pi} \evec_i \frac{1}{r}\suml^\infty_{n=0}
\frac{(-1)^n}{n!} \partial_{i_{1}} \cdots
\partial_{i_{n}}\mu_{{i_1} \cdots {i_n}i}(t-r/c)
\eeqan
and considering :
$$
\partial_{i_1} \cdots \partial_{i_n} f(t-r/c)= \frac{(-1)^n}{c^n}\nu_{i_1}
  \ldots \nu_{i_n} \frac{\rmd^n}{\rmd t^n}f(t-r/c)+{\it O}(1/r)
$$
one obtains the following expression for the part of $\Avec$ contributing to the
radiation:
\begin{eqnarray}\label{expArad}
\Avec_{rad}(\rvec,t)&=& \frac{\mu_0}{4\pi} \frac{1}{r} \evec_i\suml^\infty_{n=0} \frac{1}{n!c^n}\nu_{i_1} \ldots \nu_{i_n}
  \frac{\rmd^n}{\rmd t^n}\mu_{i_1 \cdots i_{n},i}(t-r/c) \nonumber\\
&=&\frac{\mu_0}{4\pi} \frac{1}{r}\sum_{n \geq 0} \frac{1}{n!c^n}
[\nuvec^{n}||\mu^{(n+1)}_{,n}]
\end{eqnarray}
with the notation:
\begin{displaymath}
\mu^{(n+1)}_{,n}=\frac{\rmd^n}{\rmd t^n}\mu^{(n+1)}
\end{displaymath}
So, considering the angular distribution of the radiation given by equation \eref{angd}  and
applying the expansion of $\Avec_{rad}$, one gets:
\begin{equation}
{\mathcal J}(\nuvec)=\frac{1}{16\pi^2\eps_0c^3}\left[
  \nuvec \times \sum\limits_{n\geq 0}\frac{1}{n!c^n}
  \left[\nuvec^n\vert\vert\mu^{(n+1)}_{,n+1} \right]\right]^2
\end{equation}
Finally the following result for the angular distribution of the radiation is obtained:
\beqan
\fl&~& 16\pi^2\eps_0 c^3{\mathcal
  J}(\nuvec)=\suml_{n\geq 0,m\geq 0} \frac{1}{n!m!c^{n+m}}
 \left[ \eps_{ijk}\eps_{ij'k'}\nu_j\nu_{j'}\nu_{i_1}\dots \nu_{i_n}
\nu_{j_1}\dots \nu_{j_m}\frac{\rmd^{n+1}}{\rmd t^{n+1}}\mu_{i_1\dots i_n,k}\right.\\
\fl&~&\left.\times\frac{\rmd^{m+1}}{\rmd t^{m+1}}
\mu_{j_1\dots j_m,k'}\right]\nonumber\\
\fl&~&=\sum\limits_{n\geq 0,m\geq 0}\frac{1}{n!m!c^{n+m}}
 \left[\left(\nuvec^n\vert\vert\mu^{(n+1)}_{,n+1}\right)_k
\left(\nuvec^m\vert\vert\mu^{(m+1)}_{,m+1}\right)_k-
\left(\nuvec^{n+1}\vert\vert\mu^{(n+1)}_{,n+1}\right)
\left(\nuvec^{m+1}\vert\vert\mu^{(m+1)}_{,m+1}\right)\right]\nonumber
\eeqan

\par In order to calculate the total radiated power the formula 
${\mathcal J}=4\pi<{\mathcal J}(\nuvec)>$ is used,
where $<f(\nuvec)>=(1/4\pi)\int f(\nuvec)d\Omega (\nuvec)$. The result is:
\begin{eqnarray}\label{rpow1}
\fl&~&{\mathcal J}=\frac{1}{4\pi\eps_0 c^3}\big<\suml_{n\geq 0,m\geq 0}
\frac{1}{n!m!c^{n+m}}
\left[ \nu_{i_1}\dots \nu_{i_n}\nu_{i_{n+1}}\dots \nu_{i_{n+m}}
\frac{\rmd^{n+1}}{\rmd t^{n+1}}\mu_{i_1\dots i_n,i}
\frac{\rmd^{m+1}}{\rmd t^{m+1}}\mu_{i_{n+1}\dots i_{n+m},i}\right.\nonumber\\
\fl&~&-\left.\nu_{i_1}\dots \nu_{i_n}\nu_{i_{n+1}}\nu_{i_{n+2}}\dots \nu_{i_{n+m+3}}
\frac{\rmd^{n+1}}{\rmd t^{n+1}}\mu_{i_1\dots i_n,i_{n+1}}
\frac{\rmd^{m+1}}{\rmd t^{m+1}}\mu_{i_{n+2}\dots i_{n+m+3}}
\right]\big>_{\nuvec}
\end{eqnarray}
In the previous equation it is usefull an averaging formula introduced in \cite{VN,Di}
and justified in \cite{Irina}: 
\beqa\label{avform}
\fl&~&<\nu_{i_1}\dots\nu_{i_{2n-1}}\nu_{i_{2n}}>=\frac{1}{(2n+1)!!} \suml_{D(i)}\delta_{i_1i_2}\dots\delta_{i_{2n-1}i_{2n}}
\eeqa
\par When working with the expansion \eref{rpow1} of the radiated power, one must have a
strict criterion regarding the comparison of the different terms
contributions. This criterion can be easily obtained if one refers,
particularly, to monochromatic sources. If the possibility to represent any type of variation in time as a superposition
of functions corresponding to the monochromatic sources is taken into account, the conclusions will
be general. In the case of the variation characterized by the pulsation
$\omega$, the term from the expansion indexed with the pair (n,m) contributes
with an order of magnitude equal to $(d/\la)^{n+m}$ to the radiated
power. Since the expression of the radiated power comes from the values of the
field in the zone where $\lambda >> d$, it is obvious that the order of magnitude
of a term is given by the sum $n+m$.  Therefore, {\it a consistent way
of using a finite number of terms in the expansion \eref{rpow1}, terms that accurately
characterize the radiation up to a given order {\it M}, is to retain all the terms corresponding to the values $n+m$ between 0 and {\it
  M}. Apparently, in literature such a procedure was not consistently used everywhere,
  as it will be shown later on}.
\par  So, considering ${\mathcal J}^{({\it M})}$
the {\it M} order term from the expansion \eref{rpow1}, we have
$$
{\mathcal J}^{(M)}=\suml_{n+m=M}{\mathcal J}^{(n,m)}.
$$
\begin{eqnarray}\label{In,m}
\fl&~&4\pi\eps_0c^3{\mathcal J}^{(n,m)}=\nonumber\\
\fl&~&\frac{1}{n!m!c^{n+m}}
\left<\left(\nuvec^n\vert\vert\mu^{(n+1)}_{,n+1}\right)_k
\left(\nuvec^m\vert\vert\mu^{(m+1)}_{,m+1}\right)_k 
-\left(\nuvec^{n+1}\vert\vert\mu^{(n+1)}_{,n+1}\right)
\left(\nuvec^{m+1}\vert\vert\mu^{(m+1)}_{,m+1}\right) \right>.
\end{eqnarray}
For particular cases,
\beqan
\fl&~&{\mathcal J}^{(0)}={\mathcal J}^{(0,0)},\;\;
{\mathcal J}^{(1)}={\mathcal J}^{(0,1)}+{\mathcal J}^{(1,0)}=0,\\
\fl&~&{\mathcal J}^{(2)}={\mathcal J}^{(0,2)}+{\mathcal J}^{(2,0)}+
{\mathcal J}^{(1,1)},\;\;{\mathcal J}^{(3)}=0,\\
\fl&~&{\mathcal J}^{(4)}={\mathcal J}^{(2,2)}+{\mathcal J}^{(1,3)}+{\mathcal J}^{(3,1)}
+{\mathcal J}^{(0,4)}+{\mathcal J}^{(4,0)}
\eeqan
where the fact that the averaged terms with $n+m$ odd are zero is considered.
\par Below we will present the first terms of the expansion:
\begin{eqnarray}\label{Imn}
\fl&~& 4\pi\eps_0 c^3 {\mathcal J}^{(0,0)}=\left<\dot{\mu}_i \dot{\mu}_i-(\nuvec\cdot\dot{\mu})_i
(\nuvec\cdot\dot{\mu})_i\right>=\dot{\vec{\mu}}^{\,2}-<\nu_i\nu_j>\dot{\mu}_i\dot{\mu}_i=
\frac{2}{3}\dot{\vec{\mu}}^{\,2} \nonumber\\
\fl&~& 4\pi\eps_0 c^3 {\mathcal J}^{(1,1)}=\frac{1}{c^2}\left<
(\nuvec\vert\vert{\ddot{\mu}}^{(2)})_i(\nuvec\vert\vert{\ddot{\mu}}^{(2)})_i
-(\nuvec\vert\vert\hat{\ddot{\mu}}^{(2)})(\nuvec\vert\vert\hat{\ddot{\mu}}^{(2)})\right>\nonumber\\
\fl&~&=\frac{1}{c^2}\left[<\nu_i\nu_j>\ddot{\mu}_{i_1i}\ddot{\mu}_{i_1i}-
<\nu_{i_1}\nu_{i_2}\nu_{i_3}\nu_{i_4}>\ddot{\mu}_{i_1i_2}\ddot{\mu}_{i_3i_4}\right]\nonumber\\
\fl&~&=\frac{1}{15c^2}\left[4\ddot{\mu}_{ij}\ddot{\mu}_{ij}-\ddot{\mu}_{ij}\ddot{\mu}_{ji}
-\ddot{\mu}_{ii}\ddot{\mu}_{jj}\right] \nonumber\\
\fl&~& 4\pi\eps_0 c^3 {\mathcal J}^{(0,2)}= 4\pi\eps_0 c^3 {\mathcal J}^{(2,0)}=
\frac{1}{2c^2}\left<\dot{\mu}_i(\nuvec^2\vert\vert{\dot{\ddot{\mu}}}^{(3)})_i
-(\vec{\nu}\cdot\dot{\vec{\mu}})(\nuvec^2\vert\vert{\dot{\ddot{\mu}}}^{(3)})\right>\nonumber\\
\fl&~&=\frac{1}{15c^2}\left[2\dot{\mu}_i \dot{\ddot{\mu}}_{jji}
-\dot{\mu}_i\dot{\ddot{\mu}}_{ijj}\right]\nonumber\\
\fl&~& 4\pi\eps_0 c^3 {\mathcal J}^{(2,2)}=\frac{1}{4c^4}
\left<(\nuvec^2\vert\vert\hat{\dot{\ddot{\mu}}}^{(3)})_i
(\nuvec^2\vert\vert{\dot{\ddot{\mu}}}^{(3)})_i-
(\nuvec^3\vert\vert{\dot{\ddot{\mu}}}^{(3)})
(\nuvec^3\vert\vert{\dot{\ddot{\mu}}}^{(3)})\right>\nonumber\\
\fl&~&\frac{1}{4c^2}\left[<\nu_{i_1}\dots\nu_{i_4}>\dot{\ddot{\mu}}_{i_1i_2i}
\dot{\ddot{\mu}}_{i_3i_4i}-<\nu_{i_1}\dots\nu_{i_6}>\dot{\ddot{\mu}}_{i_1i_2i_3}\dot{\ddot{\mu}}_{i_4i_5i_6}\right]\nonumber\\
\fl&~&4\pi\eps_0 c^3 {\mathcal J}^{(1,3)}
=4\pi\eps_0 c^3 {\mathcal J}^{(3,1)}=
\frac{1}{6c^4}\left<(\nuvec\vert\vert{\ddot{\mu}}^{(2)})_i
(\nuvec^3\vert\vert{\ddot{\ddot{\mu}}}^{(4)})_i-
(\nuvec^2\vert\vert{\ddot{\mu}}^{(2)})
(\nuvec^4\vert\vert{\ddot{\ddot{\mu}}}^{(4)})\right>\nonumber\\
\fl&~&=\frac{1}{6c^4}\left[<\nu_{i_1}\dots\nu_{i_4}>\ddot{\mu}_{i_1i}\ddot{\ddot{\mu}}
_{i_2i_3i_4i}
-<\nu_{i_1}\dots\nu_{i_6}>\ddot{\mu}_{i_1i_2}\ddot{\ddot{\mu}}_{i_3i_4i_5i_6}\right]\nonumber\\
\fl&~&4\pi\eps_0 c^3 {\mathcal J}^{(0,4)}=4\pi\eps_0 c^3 {\mathcal J}^{(4,0)}=
\frac{1}{24c^4}\left<\dot{\mu}_i(\nu^4\vert\vert{\ddot{\tdot{\mu}}}^{(5)})_i
-(\vec{\nu}\cdot\dot{\vec{\mu}})(\nu^5\vert\vert{\ddot{\tdot{\mu}}}^{(5)})\right>\nonumber\\
\fl&~&=\frac{1}{24c^4}\left[<\nu_{i_1}\dots\nu_{i_4}>\dot{\mu}_i\ddot{\tdot{\mu}}_{i_1i_2i_3i_4i}-
<\nu_{i_1}\dots\nu_{i_6}>\dot{\mu}_{i_1}\ddot{\tdot{\mu}}_{i_2i_3i_4i_5i_6}\right]
\end{eqnarray}
For ${\mathcal J}^{(2,2)},\,{\mathcal J}^{(1,3)},\,{\mathcal J}^{(0,4)}$, 
the results of the contractions with the $\delta-$ tensors will be given bellow by a simpler 
method.
\par The expressions for ${\mathcal J}^{(0,0)}, {\mathcal J}^{(1,1)},
 {\mathcal J}^{(0,2)},
{\mathcal J}^{(2,0)}$ are given by Bellotti and Bornatici in \cite{bel}. They go
further, introducing the reduced multipole moments and finding a new term, as we will see in 
the following.
\par Sometimes it is easier to use the magnetic moments defined as:
\begin{equation}\label{Mnou}
\fl\msf^{(n)}=\frac{n}{n+1}\intl_{\mathcal D}\xivec^n\times\jvec\rmd^3\xi:
 \msf_{i_1\dots i_n}=\frac{n}{n+1}\intl_{\mathcal D}\xi_{i_1}\dots\xi_{i_{n-1}}(\xivec\times\jvec)_{i_n}
\rmd^3\xi
\end{equation}
It can be shown that instead of the expansion \eref{expA1} one can use an expansion
obtained from this one by performing the substitution:
\begin{equation}\label{smiu}
\mu_{i_1\dots i_n}\longrightarrow -\eps_{i_ni_{n-1}k}\msf_{i_1\dots i_{n-2}k}
 +\frac{1}{n}\dot{\psf}_{i_1\dots i_n}
\end{equation}
The result for the vector potential is given in \cite{Vr,vr}:
\begin{eqnarray}
\Avec(\rvec,t)&=&\frac{\mu_0}{4\pi}\nablav\times
 \suml^\infty_{n=1}\frac{(-1)^{n-1}}{n!} \nablav^{n-1} \vert\vert\left[\frac{1}{r}\msf^{(n)}(t-r/c)\right]\nonumber\\
&~&+\frac{\mu_0}{4\pi}
 \suml^\infty_{n=1}\frac{(-1)^{n-1}}{n!}\nablav^{n-1}\vert\vert\left[\frac{1}{r}\dot{\psf}^{(n)}(t-r/c)\right]
\end{eqnarray}
The transformation \eref{smiu} is used in the equations \eref{Imn}:
\beqan
&~&4\pi\eps_0 c^3 {\mathcal J}^{(0,0)}=\dot{\mu}_i\dot{\mu}_i-<\nu_{i_1}
  \nu_{i_2}>\dot{\mu}_{i_1}\dot{\mu}_{i_2}=\frac{2}{3}\ddot{\vec{p}}^{\,2}\nonumber\\
&~&4\pi\eps_0 c^3 {\mathcal J}^{(1,1)}=\frac{1}{c^2}\left[<\nu_{i_1}\nu_{i_2}>
 (-\eps_{ii_1k}\ddot{m}_k+\frac{1}{2}\tdot{\psf}_{i_1i})
 (-\eps_{ii_2k'}\ddot{m}_{k'}+\frac{1}{2}\tdot{\psf}_{i_2i}) \right.\nonumber\\
&~&-\left.<\nu_{i_1}\nu_{i_2}\nu_{i_3}\nu_{i_4}>
 (-\eps_{i_2i_1k}\ddot{m}_k+\frac{1}{2}\tdot{\psf}_{i_1i_2})
 (-\eps_{i_4i_3k'}\ddot{m}_{k'}+\frac{1}{2}\tdot{\psf}_{i_3i_4})\right]\nonumber\\
&~&=\frac{1}{c^2}\left[<\nu_{i_1}\nu_{i_2}>\left(\eps_{ii_1k}\eps_{ii_2k'}
 \ddot{m}_k\ddot{m}_{k'}-\frac{1}{2}\eps_{ii_1k}\ddot{m}_k\tdot{\psf}_{i_2i}
 -\frac{1}{2}\eps_{ii_2k}\ddot{m}_k\tdot{\psf}_{i_1i}\right.\right.\nonumber\\
&~&\left.\left.+\frac{1}{4}\tdot{\psf}_{i_1i}
 \tdot{\psf}_{i_2i}\right)
-\frac{1}{4}<\nu_{i_1}\dots\nu_{i_4}>\tdot{\psf}_{i_1i_2}\tdot{\psf}_{i_3i_4}
  \right]
\eeqan
Here the fact that the contractions of symmetric and
antisymmetric pairs of indices cancel is considered. Because $<\nu_i\nu_j>=(1/3)\delta_{ij}$, $<\nu_{i_1}\nu_{i_2}>\eps_{ii_1k}\ddot{m}_k\tdot{\psf}_{i_2i}=(1/3)\eps_{ii_1k} \ddot{m}_k\tdot{\psf}_{i_1i}=0$ {\it etc} and $\eps_{ii_1k}\eps_{ii_1l}=
 2\delta_{kl}$, so what is left is:
\beqan
&~& 4\pi\eps_0 c^3 {\mathcal J}^{(1,1)}=\frac{2}{3}\ddot{\vec{m}}^2+\frac{1}{20c^2}\tdot{\psf}_{ij}\tdot{\psf}_{ij}-\frac{1}{60c^2}\tdot{\psf}_{ii}\tdot{\psf}_{jj}\nonumber\\
&~&4\pi\eps_0 c^3 {\mathcal J}^{(0,2)}=4\pi\eps_0 c^3 {\mathcal J}^{(2,0)}
=\frac{1}{2c^2}\left<\dot{\mu}_i(\nuvec^2\vert\vert\hat{\tdot{\mu}}^{(3)})_i-
 (\vec{\nu}\cdot\dot{\vec{\mu}})(\nuvec^3\vert\vert\hat{\tdot{\mu}}^{(3)})\right> \nonumber\\
&~&=\frac{1}{2c^2}\left[<\nu_{i_1}\nu_{i_2}>\dot{\mu}_i\tdot{\mu}_{{i_1} {i_2} {i}}-
 <\nu_{i_1}\dots\nu_{i_4}>\dot{\mu}_{i_1}\tdot{\mu}_{{i_2} {i_3} {i_4}}\right]\nonumber\\
 &~&=\frac{1}{2c^2}\left[<\nu_{i_1}\nu_{i_2}>\ddot{p}_i(-\eps_{{i} {i_2} {k}}\tdot{\msf}_
 {i_1 k}+\frac{1}{3}\dot{\tdot{\psf}}_{{i_1} {i_2} {i}})
- <\nu_{i_1}\dots\nu_{i_4}>\ddot{p}_{i_1}\right. \nonumber\\
&~&\left.(-\eps_{i_4 i_3 k}\tdot{\msf}_{i_2k}+\frac{1}{3}\dot{\tdot{\psf}}_{i_2 i_3 i_4})\right]
 =\frac{1}{2c^2}\big[-\frac{1}{3}\ddot{p}_i\tdot{\Nsf}_i+\frac{2}{45}\ddot{p}_i
 \dot{\tdot{\psf}}_{iqq}\big]
\eeqan
where
\beq\label{Ni}
 \fl\Nsf_i=\eps_{ips}\msf_{ps}=\frac{2}{3}\intl_{\mathcal D}[\xivec\times(\xivec\times
 \jvec)]\rmd^3\xi
\eeq
\beqan
\fl&~&4\pi\eps_0 c^3 {\mathcal J}^{(2,2)}=\frac{1}{4c^4}\left[<\nu_{i_1}\dots \nu_{i_4}>
\tdot{\mu}_{i_1i_2i}\tdot{\mu}_{i_3i_4i}-<\nu_{i_1}\dots \nu_{i_6}>
\tdot{\mu}_{i_1i_2i_3}\tdot{\mu}_{i_4i_5i_6}\right]\nonumber\\
\fl&~&=\frac{1}{4c^4}\left[<\nu_{i_1}\dots \nu_{i_4}>\left(\eps_{ii_2k}\eps_{ii_4k'}
\tdot{\msf}_{i_1k} \tdot{\msf}_{i_3k'}-\frac{2}{3}\eps_{ii_2k} \tdot{\msf}_{i_1k}
 \ddot {\ddot{ \psf}}_{i_3i_4i}\right.\right.\\
\fl&~&\left.\left.+
\frac{1}{9} \ddot{\ddot{\psf}}_{i_1i_2i}\ddot{\ddot{\psf}}_{i_3i_4i}\right)
-\frac{1}{9}<\nu_{i_1}\dots \nu_{i_6}> \ddot{\ddot{\psf}}_{i_1i_2i_3} 
\ddot{\ddot{\psf}}_{i_4i_5i_6}\right]\nonumber\\
\fl&~&=\frac{1}{4c^4}\left[\frac{1}{15}(4\tdot{\msf}_{ij}\tdot{\msf}_{ij}
-\tdot{\msf}_{ij}\tdot{\msf}_{ji})
+\frac{1}{9\times 15}(\ddot{\ddot{\psf}}_{qqi} \ddot{\ddot{\psf}}_{qqi}+
2\ddot{\ddot{\psf}}_{ijk}\ddot{\ddot{\psf}}_{ijk})-\frac{2}{45}
\Nsf_i\ddot{\ddot{\psf}}_{qqi}\right.\nonumber\\
\fl&~&-\left.\frac{1}{105}(\ddot{\ddot{\psf}}_{qqi}\ddot{\ddot{\psf}}_{qqi}+
\frac{2}{3}\ddot{\ddot{\psf}}_{ijk}\ddot{\ddot{\psf}}_{ijk})\right]\nonumber\\
\fl&~&=\frac{1}{4c^4}\left[\frac{1}{15}(4\tdot{\msf}_{ij}\tdot{\msf}_{ij}-\tdot{\msf}_{ij}
\tdot{\msf}_{ji})
-\frac{2}{45}\Nsf_i\ddot{\ddot{\psf}}_{qqi}\right.
\left.-\frac{2}{9\times 105}\ddot{\ddot{\psf}}_{qqi}\ddot{\ddot{\psf}}_{qqi}+
\frac{8}{9\times 105}
\ddot{\ddot{\psf}}_{ijk} \ddot{\ddot{\psf}}_{ijk}\right]
\eeqan
\beqan
\fl&~&4\pi\eps_0c^3{\mathcal J}^{(1,3)}=4\pi\eps_0{\mathcal J}^{(3,1)}\\
\fl&~&=\frac{1}{6c^4}\left[<\nu_{i_1}\dots \nu_{i_4}>(\eps_{ii_1k}\eps_{ii_4k'}
\ddot{\msf}_k\dot{\tdot\msf}_{i_2i_3}-\frac{1}{4}\eps_{ii_1k}\ddot{\msf}_k
\ddot{\tdot\psf}_{i_2i_3i_4i}\right.\\
\fl&~&-\left.\frac{1}{2}\eps_{ii_4k}\dot{\tdot\msf}_{i_2i_3k}\tdot{\psf}_{i_1i}+
\frac{1}{8}\tdot{\psf}_{i_1i}
\ddot{\tdot\psf}_{i_2i_3i_4i})-\frac{1}{8}<\nu_{i_1}\dots \nu_{i_6}>\tdot{\psf}_{i_1i_2}
\ddot{\tdot\psf}_{i_3i_4i_5i_6}\right]\\
\fl&~&=\frac{1}{6c^4}\left[\frac{4}{15}\ddot{\msf}_k\dot{\tdot\msf}_{qqk}-\frac{1}{15}
\Nsf^{(3,1)}_{li}\tdot{\psf}_{li}
-\frac{3}{8\times 105}\tdot{\psf}_{qq}\ddot{\tdot\psf}_{iijj}+
\frac{9}{8\times 105}\tdot{\psf}_{ij}\ddot{\tdot{\psf}}_{qqij}\right]
\eeqan
Here, again, there is a notation 
\beq\label{Nij}
\Nsf^{(3,1)}_{ij}=\eps_{jps}\msf_{ips}.
\eeq
\beqan
\fl&~&4\pi\eps_0c^3{\mathcal J}^{(0,4)}=4\pi\eps_0c^3{\mathcal J}^{(4,0)}\\
\fl&~&=\frac{1}{24c^4}\left[<\nu_{i_1}\dots \nu_{i_4}>\ddot{p}_i(-\eps_{ii_4k}
\ddot{\tdot{\msf}}_{i_1i_2i_3k}+\frac{1}{5}\tdot{\tdot{\psf}}_{i_1i_2i_3i_4i})\right.\\
\fl&~&\left.-<\nu_{i_1}\dots \nu_{i_6}>\ddot{p}_{i_1}(-\eps_{ii_4k}
\ddot{\tdot{\msf}}_{i_1i_2i_3k}+\frac{1}{5}\tdot{\tdot{\psf}}_{i_2i_3i_4i_5i_6})\right]
\eeqan
The terms where the symmetric-antisymmetric tensor contractions are present cancel again,
 giving the final result for ${\mathcal J}^{(0,4)}$:
$$
 4\pi\eps_0c^3{\mathcal J}^{(0,4)}=4\pi\eps_0c^3{\mathcal J}^{(4,0)}=\frac{1}{24c^4}\left[-\frac{1}{5}\eps_{ijk}
 \ddot{\tdot{\msf}}_{qqjk}\ddot{p}_i+\frac{2}{5\times 35}\ddot{p}_i\tdot{\tdot{\psf}}_{qqjji}
 \right]
 $$


\section{Symmetrising and "Detracing" the Tensors}

\par In this section the symmetrisation and detracing method
specific for the electromagnetic moments is presented. The procedure and notations from
\cite{Vr} and \cite{sc} are used, as well as a theorem introduced by
Applequist in \cite{ap}. The first step is to symmetrise the magnetic tensors
$\msf^{(n)}$, already symmetric in the first $n-1$ indices:
\beqan
\msf_{{\scriptstyle (sym)}i_1\dots i_n}&=&\frac{1}{n}[\msf_{i_1\dots i_n}+
 \msf_{i_n\dots i_1}+\msf_{i_1,i_n\dots i_2}+\dots \msf_{i_1\dots i_n,i_{n-1}}]\nonumber\\
&\equiv& \suml_{D(i)}\msf_{i_1\dots i_n}
 \eeqan
where $\sum\limits_{D(i)}$ represents the sum over the independent
terms only, from all the permutations of the n indices. It can also be written
as:
$$
\msf_{{\scriptstyle (sym)}i_1\dots i_n}=\msf_{i_1\dots
  i_n}-\frac{1}{n}\suml^{n-1}_{\la=1}\eps_{i_\lambda
  i_nq}\Nsf^{(n,1)(\la)}_{i_1\dots i_{n-1}q}
$$
where $T^{...(\la)}_{i_1\dots }$ is the component which does not have the
$i_\lambda$ index, while:
$$
\Nsf^{(n,1)}_{i_1\dots i_{n-1}}=\eps_{i_{n-1}ps}M_{i_1\dots i_{n-2}ps}
$$
is a $n-1$ order tensor.
Further, the correspondence:
$$
\Tsf^{(n)}\rightarrow {\mathcal N}[\Tsf^{(n)}]
$$
is introduced, where ${\mathcal N}[\Tsf^{(n)}]$ is a $n-1$ order tensor:
$${\mathcal N}[\Tsf^{(n)}]_{i_1\dots i_{n-1}}=\eps_{i_{n-1}ps}\Tsf_{i_1\dots
  i_{n-2}ps}
$$
and ${\mathcal N}^k[\msf^{(n)}]\equiv \Nsf^{(n,k)}$ is a tensor of rank
$n-k$. Here are some examples:
$$
\Nsf^{(n,1)}_{i_1\dots i_{n-1}}=\frac{n}{n+1}\int\xi_{i_1}\dots \xi_{i_{n-2}}
[\xivec\times (\xivec \times\jvec)]_{i_{n-1}}\rmd^3\xi
$$
$$
\Nsf^{(n,2)}_{i_1\dots i_{n-1}}=-\frac{n}{n+1}\int\xi^2\xi_{i_1}\dots
\xi_{i_{n-3}} (\xivec \times\jvec)_{i_{n-2}}\rmd^3\xi
$$
 The following relations are also to be considered:
 \beqan
 \fl&~&{\mathcal N}^{2k}[\msf^{(n)}]\equiv \Nsf^{(n;2k)}=
 \frac{(-1)^k n}{n+1}\intl_{\mathcal D}\xi^{2k}\xivec^{n-2k}\times\jvec\rmd^3\xi=
(-1)^k\msf^{(n;k)},\\
\fl&~& {\mathcal  N}^{2k+1}[\msf^{(n)}]\equiv\Nsf^{(n;2k+1)}=\frac{(-1)^kn}{n+1}
\intl_{\mathcal D}\xi^{2k}\xivec^{n-2k-1}\times 
(\xivec\times\jvec)\rmd^3\xi=(-1)^k\Nsf^{((n,1);k)},\\
\fl&~&\;k=0,1,\dots
\eeqan
 where
$\Tsf^{(n...);k}$ is the tensor obtained from the contraction of {\it k} pairs
of indices. Particularly, the tensors introduced by equations \eref{Ni} and \eref{Nij} 
are 
$$\Nsf_i=\Nsf^{(2,1)}_i,\;\;\Nsf_{ij}=\Nsf^{(3,1)}_{ij}.$$
\par For the reduction of a totally symmetric tensor to a traceless
  one we write as in \cite{sc,Vr}:
$$
\widetilde{S}_{i_1\dots i_n}=S_{i_1\dots i_n}-\suml_{D(i)}\delta_{i_1i_2}\Lambda[S^{(n)}]_{i_3,i_4\dots i_n}
$$
where {\it S} is a totally symmetric tensor and $\Lambda[S^{(n)}]$ is  a totally 
symmetric tensor of rank $n-2$. Further the following notations are used:
$$\Lambda[\msf^{(n)}]=\Lambda^{(n-2)},\;\; \Lambda[\psf^{(n)}]=\Pi^{(n-2)}.$$
Moreover, sometimes, for writing  some equations in a simpler form, we will also use  the notation 
$\Lambda[\Tsf^{(n)}]$ for $\Tsf^{(n)}$ an arbitrary tensor but by this notation we will 
suppose that the symmetrization of $\Tsf^{(n)}$ is implied i.e.
$$\Lambda[\Tsf^{(n)}]\equiv\Lambda[\Tsf^{(n)}_{\scriptstyle sym}] .$$

\par In \cite{ap} a general procedure of detracing a symmetric tensor is
presented. In our case the result of this procedure may be written as:
\beqan
\Lambda[S^{(n)}]_{i_3\dots
  i_n}&=&\suml^{[n/2]}_{m=1}\frac{(-1)^{m-1}(2n-1-2m)!!} {(2n-1)!!m}\times \nonumber\\
&\times&\suml_{D(i)}\delta_{i_3\,i_4}\dots\delta_{i_{2m-1}\,i_{2m}}S^{n;m}_{i_{2m+1}
  \dots i_n}
\eeqan
Using the definitions and notations presented above it is easy to obtain the
following useful results:
\begin{itemize}
\item detracing the electric moment tensors:
\begin{equation}
\Pi=\Lambda[\psf^{(2)}]:\;\;\Pi=\frac{1}{3}\psf_{ii}
\end{equation}
\begin{equation}
\Pi^{(1)}=\Lambda[\psf^{(3)}]:\;\;\Pi_i=\frac{1}{5}\psf_{qqi}
\end{equation}
\begin{equation}
\Pi^{(2)}=\Lambda[\psf^{(4)}]:\;\;\Pi_{ij}=\frac{1}{7}\psf_{qqij}-\frac{1}{70}
\psf_{qqll}\delta_{ij}
\end{equation}
\begin{equation}
\Pi^{(3)}=\Lambda[\psf^{(5)}]:\;\;\Pi_{ijk}=\frac{1}{11}\psf_{qqijk}-\frac{1}{11\times
  18}\suml_{D(i)}\delta_{ij}\psf_{qqllk}
\end{equation}
\item detracing the magnetic moments tensors:
\begin{equation}
\Lambda=\Lambda[\msf^{(2)}_{\scriptstyle sym}]=0
\end{equation}
\begin{equation}
\Lambda^{(1)}=\Lambda[\msf^{(3)}_{\scriptstyle
  sym}]:\;\;\Lambda_i=\frac{1}{15}\msf_{qqi}
\end{equation}
\begin{equation}
\Lambda^{(2)}=\Lambda[\msf^{(4)}_{\scriptstyle sym}]:\;\;
 \Lambda_{ij}=\frac{1}{28}(\msf_{qqij}+\msf_{qqji})
\end{equation}
\begin{equation}
\Lambda^{(3)}=\Lambda[\msf^{(5)}_{\scriptstyle sym}]:\;\;
 \Lambda_{ijk}=\frac{1}{45}\suml_{D}\msf_{qqijk}-\frac{1}{14\times 45}\suml_{D}
 \delta_{ij}\msf_{qqllk}
\end{equation}
\end{itemize}

\section{The Total Radiated Power in Terms of Reduced Moments}

\par The usual procedure is to emphasize, in the expression of ${\mathcal J}$, the
reduced tensors by utilizing the reduction relations given in the previous
section. The well-known notations for dipole moments are used:
$$
\psf_i=p_i:\;\vec{p},\;\;\;\;\msf_i=m_i:\;\vec{m}
$$
The static expressions of the reduced tensors of electric and magnetic polarizations
 are ${\mathcal  P}^{(n)}$ and ${\mathcal M}^{(n)}$  with :
$$
{\mathcal P}_{i_1\dots i_n}=\frac{(-1)^n}{(2n-1)!!}\intl_{\mathcal D} \rho(\rvec,t) r^{2n+1}\nablav^n\frac{1}{r}\rmd^3x
$$
and  \cite{sc},
$$
{\mathcal M}_{i_1\dots i_n}=\frac{(-1)^n}{(n+1)(2n-1)!!}\suml^n_{\la=1}
 \intl_{\mathcal
 D}r^{2n+1}[\jvec(\rvec,t)\times\nablav]_{i_{\la}}\d^{(\la)}_{i_1\dots i_n}
 \frac{1}{r}\rmd^3x
$$
Below there are the results of the reduction of the first terms in the
expansion of ${\mathcal J}$, which correspond to the limitation to the order
$(d/\lambda)^4$:
\beqa\label{Imnr}
\fl&~&4\pi\eps_0c^3{\mathcal J}^{(0,0)}=\frac{2}{3}\ddot{\vec{p}}^{\,\,2}\nonumber\\
\fl&~&4\pi\eps_0c^3({\mathcal J}^{(0,2)}+{\mathcal
  I}^{(2,0)})=-\frac{4}{3c^2}\ddot{p}_i \tdot{\Tsf}_i,\;\; \Tsf_i=\frac{1}{4}\Nsf_i-\frac{1}{6}\dot{\Pi}_i\nonumber\\
\fl&~&4\pi\eps_0c^3{\mathcal
  I}^{(1,1)}=\frac{2}{3c^2}\ddot{\vec{m}}^{\,2}+\frac{1}{20c^2} \tdot{\mathcal P}_{ij}\tdot{\mathcal P}_{ij}\nonumber\\
\fl&~& 4\pi\eps_0c^3{\mathcal J}^{(2,2)}=\frac{1}{4c^4}\left[\frac{1}{5}
 \tdot{\mathcal M}_{ij}\tdot{\mathcal M}_{ij}+\frac{1}{6}\tdot{\Nsf}_i\tdot{\Nsf}_i -\frac{2}{9}\tdot{\Nsf}_i\dot{\tdot{\Pi}}_i+\frac{2}{27}\dot{\tdot{\Pi}}_i
 \dot{\tdot{\Pi}}_i+\frac{8}{945}\dot{\tdot{\mathcal P}}_{ijk}\dot{\tdot{\mathcal P}}_{ijk}
 \right]\nonumber\\
\fl&~&4\pi\eps_0c^3({\mathcal J}^{(1,3)}+{\mathcal J}^{(3,1)})=
\frac{1}{3c^4}\left[
\frac{4}{15}\ddot{m}_k\ddot{\ddot{\msf}}_{qqk}-\frac{1}{15}\ddot{\ddot{\Nsf}}^{(3,1)}_{ij}
\tdot{\mathcal P}_{ij}
+\frac{3}{40}\ddot{\tdot{\Pi}}_{ij}\tdot{\mathcal P}_{ij}\right]
\nonumber\\
\fl&~&4\pi\eps_0c^3({\mathcal J}^{(0,4)}+{\mathcal J}^{(4,0)})=
\frac{1}{12c^4}\left[-\frac{1}{5}\ddot{p}_i\ddot{\tdot{\Nsf}}^{(4,1)}_{qqi}+\frac{4}{25}\ddot{p}_i
\tdot{\tdot{\Pi}}_{jji}\right]
\eeqa
The  toroid dipole moment \cite{dub1,dub2,dub3} was introduced, with the definition:
$$
\Tsf_i=\frac{1}{4}\Nsf_i-\frac{1}{6}\dot{\Pi}_i
$$
There was also used the simplified notation $\Nsf_i=\Nsf^{(2,1)}_i$, while
$\widetilde{\Pi}^{(2)}$ is a traceless tensor (that is $\Pi^{(2)}$
``detraced''):
$$
\widetilde{\Pi}_{ik}=\Pi_{ik}-\frac{1}{3}\Pi_{qq}\delta_{ik}=
 \frac{1}{7}\psf_{qqik}-\frac{1}{21}\psf_{qqll}\delta_{ik}
$$
The sum of terms from the equation \eref{Imnr} represents the total radiated power
expanded until the fourth order with respect to $d/\lambda$, that is:
 $$\suml_{n+m\leq 4}{\mathcal J}^{(n,m)}$$
It is verified the relation:
$$
\frac{2}{3}\ddot{\vec{p}}^2-\frac{4}{3c^2}\ddot{p}_i\tdot{\Tsf}_i+\frac{1}{24c^4} \tdot{\Nsf}_i\tdot{\Nsf}_i-\frac{1}{18c^4}\tdot{\Nsf}_i\dot{\tdot{\Pi}}_i
 +\frac{1}{54} \dot{\tdot{\Pi}}_i\dot{\tdot{\Pi}}_i=\frac{2}{3}\left(\ddot{\vec{p}}- \frac{1}{c^2}\tdot{\vec{\Tsf}}\right)^2
$$
By eliminating the higher order derivatives, we may write:
$$
\frac{1}{15}\Nsf_{ik}{\mathcal P}_{ik}-\frac{3}{40}\dot{\Pi}_{ik} {\mathcal P}_{ik}=\frac{1}{15}\widetilde{\Nsf}_{ik}{\mathcal P}_{ik}-\frac{3}{40}
 \dot{\widetilde{\Pi}}_{ik} {\mathcal P}_{ik}=\frac{3}{10}\left(\frac{1}{9}
 \widetilde{\Nsf}_{ik}-\frac{1}{4}\dot{\widetilde{\Pi}}_{ik}\right){\mathcal
 P}_{ik}
$$
with the notation: $\Nsf_{ik}=\Nsf^{(3,1)}_{ik}$.
The so-called quadrupole toroid moment \cite{dub1,dub2,dub3,po} is also defined by:
$$
\Tsf_{ik}=\frac{1}{9}\widetilde{\Nsf}_{ik}-\frac{1}{4}\dot{\widetilde{\Pi}}_{ik}
$$
Explicitly, the two toroid moments are written as:
\beqan
\Tsf_i&=&\frac{1}{6}\intl_{\mathcal D}[\xivec\times(\xivec\times\jvec)]_i
 \rmd^3\xi-\frac{1}{30}\dot{\psf}_{qqi}\nonumber\\
&=& \frac{1}{6}\intl_{\mathcal D}[(\xivec\cdot\jvec)\xi_i-\xi^2j_i]\rmd^3\xi
 -\frac{1}{30}\intl_{\mathcal D}\xi^2\xi_i\dot{\rho}\rmd^3\xi
\eeqan
but since:
$$\intl_{\mathcal D}\xi^2\xi_i\dot{\rho}\rmd^3\xi=-\intl_{\mathcal D}
 \xi^2\xi_i\nablav\cdot\jvec\rmd^3\xi=\intl_{\mathcal D}\jvec\cdot(\xi^2\xi_i)\rmd^3\xi=
 \intl_{\mathcal D}[2(\xivec\cdot\jvec)\xi_i+\xi^2j_i]\rmd^3\xi$$
it results:
$$
\Tsf_i=\frac{1}{10}\intl_{\mathcal
  D}\big[(\xivec\cdot\jvec)\xi_i-2\xi^2j_i\big] \rmd^3\xi
$$
In a similar way:
$$
\Tsf_{ik}=\frac{1}{42}\intl_{\mathcal D}\big[4(\xivec\cdot\jvec)\xi_i\xi_k-
 5\xi^2(\xi_ij_k+\xi_kj_i)+2\xi^2(\xivec\cdot\jvec)\delta_{ik}\big]\rmd^3\xi
$$
After regrouping the terms and considering the above presented relations, one
gets:
\beqa\label{IBB}
\fl&~&4\pi\eps_0 c^3{\mathcal J}^{(4)}=\frac{2}{3}\left(\ddot{\vec{p}}-\frac{1}{c^2}\tdot{\vec{\Tsf}}
 \right)^2+\frac{2}{3c^2}\ddot{\vec{m}}^{\;2}+\frac{1}{20c^2}
 \tdot{\mathcal P}_{ij}\tdot{\mathcal P}_{ij}-\frac{1}{10c^4}\ddot{\ddot{\Tsf}}_{ij}
 \tdot{\mathcal P}_{ij}\nonumber\\
\fl&~&+\frac{4}{45c^4}\ddot{m}_k \ddot{\ddot{\msf}}_{qqk}-\frac{1}{60c^4}\ddot{p}_i\left(\ddot{\tdot{\Nsf}}^{(4,1)}_{qqi}-
 \frac{4}{5}\tdot{\tdot{\Pi}}_{qqi}\right)
+\frac{1}{20c^4}\tdot{\mathcal M}_{ij}\tdot{\mathcal M}_{ij}+\frac{2}{945c^4}
 \ddot{\ddot{\mathcal P}}_{ijk}\ddot{\ddot{\mathcal P}}_{ijk}
\eeqa
If the expansion of the radiated power up to the second order with
respect to $d/\lambda$ is considered, only the following sum will be kept:
\beqan
{\mathcal J}^{(2)}&=&{\mathcal J}^{(0,0)}+{\mathcal J}^{(1,1)}+2{\mathcal
 I}^{(0,2)}\nonumber\\
&=&\frac{1}{4\pi\eps_0c^3}\left[ \frac{2}{3}\ddot{\vec{p}}^2+
\frac{2}{3c^2}\ddot{\vec{m}}^2 -\frac{4}{3c^2}\ddot{p}_i\tdot{\Tsf}_i+
\frac{1}{20c^2}\tdot{\mathcal P}_{ij}\tdot{\mathcal P}_{ij}
 \right]
\eeqan
This is, actually, a result obtained by Belloti and Bornatici
\cite{bel}. They observe that, compared to the expression for the {\it
  dipolar magnetic-quadrupolar electric} radiation from the books of Jackson
and Landau \cite{jack,La} as well as from many other electrodynamics
books (including the one by C. Vrejoiu \cite{Vr}), this result contains a
supplementary term represented by the contribution of the vector
$\vec{T}$.
 We point out that in \cite{jack,La} the goal is only to calculate the isolated 
 contributions 
 of the electric dipole, electric  4-pole and magnetic dipole to the total radiated 
 power without regarding it as a result of an expansion.  
 If one associates a multipole moment to an elementary system, one obtains an 
 isolated contribution of this multipole, but when one considers a composite system, 
 all multipoles, giving contributions of the same order of magnitude, must be considered. 
 This is the reason for which 
 the result from \cite{jack,La} should not to be considered erroneous. 

 Without knowing, probably, the results published ever since 1974 by
Dubovik et al. \cite{dub1} and all the other publications refering to toroid
moments, they consider this term as being something new. On the other hand,
Bellotti and Bornatici give a correct result, if one considers how far they go
with the expansion of the radiated power.
\par The impression is that in general in literature similar results contain 
a series of confusions due to the inconsistent application 
of the criteria according to which different terms of this expansion
must be compared. For instance, in \cite{dub3} a result which is supposed to
represent the expansion including the fourth order terms is presented:
\beqa\label{dub}
4\pi\eps_0c^3{\mathcal J}_{\scriptstyle Dubovik}&=&\frac{2}{3}\left(\ddot{\vec{p}}-
\frac{1}{c^2}\tdot{\vec{\Tsf}}
\right)^2+\frac{2}{3c^2}\ddot{\vec{m}}^{\;2}\nonumber\\
&+&\frac{1}{20c^2}
\left(\tdot{\mathcal
    P}_{ik}-\frac{1}{c^2}\tdot{\Tsf}_{ik}\right)\left(\tdot{\mathcal
    P}_{ik}-\frac{1}{c^2}\tdot{\Tsf}_{ik}\right)\nonumber\\
&+&\frac{2}{945c^4}
\left(\ddot{\ddot{\mathcal P}}_{ijk}\ddot{\ddot{\mathcal P}}_{ijk}+
\ddot{\ddot{\msf}}_{ijk}\ddot{\ddot{\msf}}_{ijk}\right)+\dots
\eeqa
Appart from the mistake (of course, typing error) that $\Tsf_{ik}$ is derived
only three times, there are certain things that must  be observed:
\begin{itemize}
\item the term ${\mathcal M}_{ijk}{\mathcal M}_{ijk}$ from this expression is
  of the sixth order (with the $1/\lambda$ criterion);
\item the term $\Tsf_{ik}\Tsf_{ik}$ is also of the sixth order;
\item these terms should not be present if in the expansion of  ${\mathcal J}$ the
  terms up to the sixth order are not consiered;
\item the term ${\mathcal M}_{ij}{\mathcal M}_{ij}$ is missing, as well as
  other terms that should be present, at least up to the fourth order;
\item the sign three points "\ldots" at the end of the expression is losing its
  usual sense when a finite number of terms is considered in a series
  expansion.
\end{itemize}
\par In some more recent papers, as for example in \cite{ra}, it is claimed that, 
 as an application of more general formulae,  
  the expansion up to the order $1/c^5$ is given (that is, exactly how it is
  considered in \cite{bel}). The $1/c$ criterion is correlated with the
  wavelength criterion if one takes into account the powers of the parameter
  $1/c$ together with the orders of the partial temporal derivatives in the
  expansion of the radiated power. The expression given in \cite{ra} (written
  with our notations)  is:
\begin{eqnarray}\label{Irv}
\fl&~&4\pi\eps_0c^3{\mathcal J}_{\scriptstyle R-V}=
\frac{2}{3}\left(\ddot{\vec{p}}-\frac{1}{c^2}\tdot{\vec{\Tsf}}\right)^2
+\frac{4}{45c^4}\ddot{m}_k\ddot{\ddot{\msf}}_{qqk}
+\frac{1}{20c^2}\tdot{{\mathcal P}}_{ij}\tdot{\mathcal P}_{ij}
+\frac{1}{20c^4}\tdot{{\mathcal M}}_{ij}\tdot{\mathcal M}_{ij}
\end{eqnarray}
where, in order to identify it with the equation (4.12) from \cite{ra}, one must consider the definitions of the quadrupol moments and the identity:
$$
\msf_{qqk}=\frac{3}{4}\int\xi^2(\xivec\times\jvec)_k\rmd^3\xi=
\frac{1}{2}\vec{\bar{\rho}}^{\;2}
$$
this last parameter being the one used in the equation \eref{Irv}.
\par It is easy to observe that if one tries to interpret equation \eref{Irv} as 
the expansion of ${\mathcal J}$ up to the second order, the fourth order term 
$\Tsf^2$ is present without justification. On the other hand, if the same equation 
is considered as an expansion up to the fourth order, many terms are missing.

\section{Reduction of the Multipole Tensors and Gauge Invariance}

\par In this section the results  from \cite{vr,Di} will be presented in a more systematic 
and concise way. A different  explanation  of these results is given 
in Appendix A using the formalism of Dirac's $\delta-$ function.
These results present the possibility of expressing the electromagnetic 
potentials, as well as the field, exclusively in terms of reduced moments, that is 
moments represented by reduced tensors (symmetric and traceless). This procedure 
allows, particularly, to express the radiated power in a very simple general form, 
as it will be shown below.
\par Let us consider the expansion of the potentials with the moments $\msf^{(n)}$ 
(see equation \eref{Mnou}) and $\psf^{(n)}$ (see equation \eref{dfi}):
\begin{eqnarray}\label{A}
\Avec(\rvec,t)&=&\frac{\mu_0}{4\pi}\nablav\times \suml^\infty_{n=1}\frac{(-1)^{n-1}}{n!}
 \nablav^{n-1} \vert\vert\left[\frac{1}{r}\msf^{(n)}(t-r/c)\right]\nonumber\\
&~&+\frac{\mu_0}{4\pi} \suml^\infty_{n=1}\frac{(-1)^{n-1}}{n!}\nablav^{n-1}\vert\vert\left[\frac{1}{r}
 \dot{\psf}^{(n)}(t-r/c)\right]
 \end{eqnarray}
\begin{equation}\label{Phi}
\Phi(\rvec,t)=\frac{1}{4\pi\eps_0}\suml_{n\geq 0}\frac{(-1)^n}{n!}\nablav^n\vert\vert
\left[\frac{\psf^{(n)}(t-r/c)}{r}\right]
\end{equation}
The basic idea is that the tensors $\msf^{(n)}$ and $\psf^{(n)}$ can be replaced by reduced tensors. This leads to a modification of some of the inferior order moments, such that finally  all the moments up to (and including) the {\it n}th order moment must be expressed by reduced tensors. Not all of them, though,
 reduce to the static expressions ${\mathcal M}$ and ${\mathcal P}$. The final results for these moments are $\widetilde{\msf}$ and $\widetilde{\psf}$. Formally, as a result of this procedure, the expressions for the
 potentials expansions up to the {\it n}th order will be given by the equations \eref{A}  
 and \eref{Phi} with the substitutions $\psf\longrightarrow \widetilde{\psf}$ and 
 $\msf\longrightarrow \widetilde{\msf}$.
\par As a basis for this procedure one could take the following observations resuming 
systematically the results from \cite{vr}:
\begin{itemize}
\item  Let $\Lsf^{(n)}$ be a $\msf^{(n)}$ -type tensor {\it i.e.}
\beq\label{defL}
\fl \Lsf\;\mbox{is\;fully\;symmetric\; in\;the\;first\; }\;n-1\;\mbox{indices},\;\;
\Lsf_{i_1\dots i_{n-1}\,i_k}=0$ if $k\leq n-1.
\eeq
 The symmetrisation of $\Lsf$ is realised with the relation:
\begin{equation}\label{defLsym}
\Lsf_{{\scriptstyle (sym)}i_1\dots i_n}=\Lsf_{i_1\dots i_n}-\frac{1}{n}
\suml^{n-1}_{\la=1}\eps_{i_\lambda i_nq}{\mathcal N}^{(\la)}_{i_1\dots i_{n-1}q}
\big[\Lsf^{(n)}\big]
\end{equation}
with:
$${\mathcal N}_{i_1\dots i_{n-1}}\big[\Lsf^{(n)}\big]=\eps_{i_{n-1}ps}
\Lsf_{i_1\dots i_{n-2}ps}$$
In this case:
\par a) the substitution:
\begin{equation}\label{subst1}
\msf_{i_1\dots i_n}\longrightarrow\msf_{(\scriptstyle L)i_1\dots i_n}=\msf_{i_1\dots i_n}-
\frac{1}{n}\suml^{n-1}_{\la=1}\eps_{i_\lambda i_nq}{\mathcal N}^{(\la)}_{i_1\dots i_{n-1}q}
\big[\Lsf^{(n)}\big]
\end{equation}
produces changes of the potentials which, up to a gauge transformation, are compensated by the following transformation:
\begin{equation}
\psf^{(n-1)}\longrightarrow \psf^{(n-1)}-\frac{n-1}{c^2n^2}\dot{\mathcal N}
\big[\Lsf^{(n)}\big];\;\;\;\;\;\;\;\;\;\;\;\;\;\;\;\;\;\;(I)
\end{equation}
\par b) the substitution:
\begin{equation}\label{subst2}
\psf_{i_1\dots i_n}\longrightarrow\psf_{({\scriptstyle L})i_1\dots i_n}-
\frac{1}{n}\suml^{n-1}_{\la=1}\eps_{i_\lambda i_nq}{\mathcal N}^{(\la)}_{i_1\dots i_{n-1}q}
\big[\Lsf^{(n)}\big]
\end{equation}
produces changes of the potentials which are compensated by the following transformation:
\begin{equation}
\msf^{(n-1)}\longrightarrow \msf^{(n-1)}+\frac{n-1}{n^2}\dot{\mathcal N}
\big[\Lsf^{(n)}\big];\;\;\;\;\;\;\;\;\;\;\;\;\;\;\;\;\;\;(II)
\end{equation}
\item Let $\Ssf^{(n)}$ be a  tensor and the detracing operation for it:
\begin{equation}\label{defS}
\widetilde{\Ssf}_{i_1\dots i_n}=\Ssf_{i_1\dots i_n}
-\suml_{D(i)}\delta_{i_1i_2}\Lambda_{i_3\dots i_n}\big[\Ssf^{(n)}\big]
\end{equation}
Then:
\par c) the substitution:
\begin{equation}\label{subst3}
\msf_{i_1\dots i_n}\longrightarrow\msf_{({\scriptstyle S})i_1\dots i_n}-
\suml_{D(i)}\delta_{i_1i_2}\Lambda_{i_3\dots i_n}\big[\Ssf^{(n)}\big]
\end{equation}
produces changes of the potentials which, up to a gauge transformation, are compensated by the following transformation:
\begin{equation}
\msf^{(n-2)}\longrightarrow\msf^{(n-2)}+\frac{n-2}{2c^2n}
\ddot{\Lambda}\big[\Ssf^{(n)}\big];\;\;\;\;\;\;\;\;\;\;\;\;\;\;(III)
\end{equation}
\par d) the substitution:
\begin{equation}
\psf_{i_1\dots i_n}\longrightarrow\psf_{({\scriptstyle S})i_1\dots i_n}-
\suml_{D(i)}\delta_{i_1i_2}\Lambda_{i_3\dots i_n}\big[\Ssf^{(n)}\big]
\end{equation}
produces changes of the potentials which, up to a gauge transformation, are compensated by the following transformation:
\begin{equation}
\psf^{(n-2)}\longrightarrow\psf^{(n-2)}+\frac{n-2}{2c^2n}\ddot{\Lambda}
\big[\Ssf^{(n)}\big].\;\;\;\;\;\;\;\;\;\;\;\;\;\; (IV)
\end{equation}
\end{itemize}
These four transformation relations of the electromagnetic potentials are sufficient
for the development of a scheme in which the replacement of the multipole moments 
tensors by symmetric and  traceless tensors is presented.
\par Such a scheme, valid for expansion up to the fourth order with respect to $d/\lambda$,
 can be found in Appendix B with results written in the Tables from Appendix C. 
 It can be easily continued for higher orders. It is possible 
 also to give  results for the reduced tensors $\widetilde{\psf}^{(n)}$ and 
 $\widetilde{\msf}^{(n)}$ for arbitrary $n$ which will be given elsewhere \cite{VS}.  

\section{ Radiated Power Expressed by Transformed Moments}

By applying the transformations \eref{smiu} to the expansion of the vector potential 
of the radiated field, one gets:
\beqan
{\Avec}_{rad}(\rvec,t) &=& \frac{\mu_0}{4\pi} \frac{1}{r}\suml^\infty_{n=1}
\frac{1}{n!c^n} [\nuvec^{(n-1)} \vert\vert \msf_{,n}^{(n)}(t-r/c)]\times \nuvec\nonumber\\
&+& \frac{\mu_{0}c}{4\pi}\frac{1}{r}\suml^\infty_{n=1}
\frac{1}{n!c^n}[\nuvec^{(n-1)} \vert\vert \psf_{,n}^{(n)}(t-r/c)]
\eeqan
The radiated power (angular distribution) will be given by:
\beqan
&~&4\pi\eps_0{\mathcal J}(\nuvec)=\nonumber\\
&~&=\suml^\infty_{n=1}\suml^\infty_{m=1}\frac{1}{n!m!c^{n+m}}\left[\left(\nuvec^{(n-1)}
 \vert\vert \msf_{,n+1}^{(n)}\right)\left(\nuvec^{(m-1)} \vert\vert \msf_{,m+1}^{(m)}
 \right)\right.\nonumber\\
&~&\left. -\left(\nuvec^{(n)} \vert\vert \msf_{,n+1}^{(n)}\right)\left(\nuvec^{(m)}
  \vert\vert \msf_{,m+1}^{(m)}\right)\right]\nonumber\\
&+&\suml^\infty_{n=1}\suml^\infty_{m=1}\frac{c^2}{n!m!c^{n+m}}\left[\left(\nuvec^{(n-1)} 
\vert\vert \psf_{,n+1}^{(n)}\right)\left(\nuvec^{(m-1)} \vert\vert \psf_{,m+1}^{(m)}
\right)\right.\nonumber\\
&~&\left.-\left(\nuvec^{(n)} \vert\vert \psf_{,n+1}^{(n)}\right)\left(\nuvec^{(m)} \vert\vert \psf_{,m+1}^{(m)}\right)\right]  \nonumber\\
&+&\suml^\infty_{n=1}\suml^\infty_{m=1}\frac{c}{n!m!c^{n+m}} \left\{ \left(\nuvec^{(n-1)} 
\vert\vert \msf_{,n+1}^{(n)}\right) \cdot \left[ \nuvec\times\left(\nuvec^{(m-1)}
 \vert\vert \psf_{,m+1}^{(m)}\right)\right] \right.\nonumber\\
&+& \left. \left(\nuvec^{(n-1)} \vert\vert \msf_{,n+1}^{(n)}\right)\cdot\left[\nuvec\times\left(\nuvec^{(m-1)} \vert\vert \psf_{,m+1}^{(m)}\right)\right] \right\}
\eeqan
Considering the procedure of reduction of the moments tensors from the expansion of 
the vector potential applied up to the $\mu$th order for the magnetic and to the 
$\eps$th order for the electric moments, 
the sum of the terms from the expansion of the radiated power which contain 
{\it exclusively} magnetic moments reduced up to $\mu$ and electric moments up to 
$\eps$ is:
\begin{eqnarray}\label{Imiu}
&~&4\pi\eps_0{\mathcal J}_{\mu,\eps}(\nuvec)=\nonumber\\
&=&\suml^\infty_{n=1}\suml^\infty_{m=1}\frac{1}{n!m!c^{n+m}}\left[\left(\nuvec^{(n-1)} 
\vert\vert \widetilde{\msf}_{,n+1}^{(n)}\right)\left(\nuvec^{(m-1)} \vert\vert 
\widetilde{\msf}_{,m+1}^{(m)}\right)\right.\nonumber\\
&~&-\left.\left(\nuvec^{(n)} \vert\vert
 \widetilde{\msf}_{,n+1}^{(n)}\right)\left(\nuvec^{(m)} \vert\vert 
 \widetilde{\msf}_{,m+1}^{(m)}\right)\right]\nonumber\\
&+&\suml^\infty_{n=1}\suml^\infty_{m=1}\frac{c^2}{n!m!c^{n+m}}\left[\left(\nuvec^{(n-1)}
 \vert\vert \widetilde{\psf}_{,n+1}^{(n)}\right)\left(\nuvec^{(m-1)} \vert\vert 
 \widetilde{\psf}_{,m+1}^{(m)}\right)\right.\nonumber\\
&~&\left. -\left(\nuvec^{(n)} \vert\vert 
 \widetilde{\psf}_{,n+1}^{(n)}\right)\left(\nuvec^{(m)} \vert\vert 
 \widetilde{\psf}_{,m+1}^{(m)}\right)\right]  \nonumber\\
&+&\suml^\infty_{n=1}\suml^\infty_{m=1}\frac{c}{n!m!c^{n+m}} \left\{ \left(\nuvec^{(n-1)}
 \vert\vert \widetilde{\msf}_{,n+1}^{(n)}\right) \cdot \left[ 
 \nuvec\times\left(\nuvec^{(m-1)} \vert\vert \widetilde{\psf}_{,m+1}^{(m)}\right)\right]
  \right.\nonumber\\
&+& \left. \left(\nuvec^{(n-1)} \vert\vert \widetilde{\msf}_{,n+1}^{(n)}\right)\cdot
 \left[\nuvec\times\left(\nuvec^{(m-1)} \vert\vert \widetilde{\psf}_{,m+1}^{(m)}\right)
  \right] \right\}
\end{eqnarray}
It is easy to understand that the above sum cannot be identified with the expansion of
 the radiated power containing the magnetic moments $\msf^{(k)}$, $k\leq\mu$ and the 
 electric moments $\psf^{(l)}$, $l\leq\eps$. This happens because some of the magnetic 
 moments $\widetilde{\msf}^{(k)}$ contain tensorial expressions built with magnetic and electric tensors of superior orders in $k$ and similar for the electric case. 
 But, once ${\mathcal J}_{\mu,\eps}$ is settled, the superior orders contributions can
  be eliminated in order to obtain the correct expansion with the $d/\lambda$ criterion.
\par Returning to the expression of the total radiated power as a function of the 
magnetic moments $\mu^{(n)}$ and considering the expansion \eref{In,m}, it results that 
this one corresponds to the expansion (\ref{Imiu}) for $\mu=M$ and $\eps=M+1$ when only the 
contributions of order not higher than $M$ are retained.
\par According to the results from \cite{Di}, expressing the total radiated power is 
easier if one uses the averaging formula \eref{avform} and the symmetric and traceless 
character of the reduced tensors.
Hence, the following properties are valid:\\
\par a) let two symmetric traceless tensors $\Atens^{(n)}$ and $\Btens^{(n)}$ and their averaged contraction:
$$
\big< \left(\nuvec^k \vert\vert \Atens^{(n)}\right) \vert\vert \left(\nuvec^{k'} \vert\vert \Btens^{(m)}\right) \big> = <\nu_{i_1}\dots \nu_{i_k}\nu_{j_1}\dots \nu_{j_k'}\Atens_{i_1\dots i_ki_{k+1}\dots i_n}\Btens_{j_1\dots j_k'j_{k'+1}\dots j_m}>
$$
This average is different from zero only for products $\delta_{i_pj_k}$ with $p=1,\dots ,k$ and $q=1,\dots ,k'$ and the following relation is valid:
$$
\big< \left(\nuvec^k \vert\vert \Atens^{(n)}\right) \vert\vert \left(\nuvec^{k'} \vert\vert \Btens^{(m)}\right) \big>= \frac{k!}{(2k+1)!!} \left[\Atens^{(n)}\vert\vert \Btens^{(m)} \right]\delta_{k'k}
$$

\par b) The terms of the last sums, containing mixed moments products, in equation 
\eref{Imiu}, give contributions of the type:
$$<\nu_{i_1}\dots \nu_{i_{n-1}}\nu_{j_1}\dots \nu_{j_{m-1}}\nu_{p}> \eps_{i_npq}\Atens_{i_1\dots i_n}\Btens_{j_1\dots j_{m-1}q}$$
but all the terms from the $\delta$ products, representing the averages of the $\nuvec$ components products, contain either $\delta_{i_kp}$ or $\delta_{pj_l}$, $k=1,\dots ,n-1$,
$l=1,\dots ,m-1$ such that, because of $\eps_{i_npq}$ and of the symmetry of $A$ and $B$, the result is zero.
Using these properties one can write the result:
\beqan
&~&{\mathcal J}_{\mu\eps}= \frac{1}{4\pi\eps_0 c^3}
\left[ \suml_{n=1}^{\mu} \frac{n+1}{nn!(2n+1)!!c^{2n}}
 \left( \widetilde{\msf}_{,n+1}^{(n)} \vert\vert \widetilde{\msf}_{,n+1}^{(n)} \right)\right.\\ 
&~&\left.+ \suml_{n=1}^{\eps} \frac{n+1}{nn!(2n+1)!!c^{2n-2}} \left( \widetilde{\psf}_{,n+1}^{(n)} 
\vert\vert \widetilde{\psf}_{,n+1}^{(n)} \right) \right]
\eeqan
\par Let us consider now different expressions of ${\mathcal J}^{(M)}$, calculated this time starting from the formula given by the equation written above. As settled, for a given value of $M$ one must consider this formula for $\mu=M$, $\eps=M+1$.
\par ${\mathcal J}^{(0)}$: $\mu=0$, $\eps=1$, only the electric dipole moment has a contribution:
$$4\pi\eps_0 c^3 {\mathcal J}^{(0)}= \frac{2}{3} \ddot{\vec{p}}^2$$
\par ${\mathcal J}^{(2)}$: $(\mu,\eps)=(2,3)$; in this case the results from the table in Appendix A are used, considering in the end only the terms corresponding to the order $M$:
$$
4\pi\eps_0 c^3 {\mathcal J}^{(2)}=\left\{ \frac{2}{3c^2} \ddot{\widetilde{\msf}}_{I}
\ddot{\widetilde{\msf}}_{I} +\frac{1}{20c^4}\tdot{\widetilde{\msf}}_{ij}
\tdot{\widetilde{\msf}}_{ij}+\frac{2}{3}\ddot{\widetilde{\psf}}_{i}
\ddot{\widetilde{\psf}}_{i}+\frac{1}{20c^2}\tdot{\widetilde{\psf}}_{ij}
\tdot{\widetilde{\psf}}_{ij} \right\}_{2}
$$
where the index of the bracket is the maximum order in $d/\lambda$ which has to be retained in this bracket. It follows that:
$$
4\pi\eps_0 c^3 {\mathcal J}^{(2)}=\left[ \frac{2}{3}\ddot{\vec{p}}^2+\frac{2}{3c^2}\ddot{\vec{m}}^2-\frac{4}{3c^2}\ddot{\vec{p}} \cdot \tdot{\vec{T}} + \frac{1}{20c^2}{\cal P}_{ij}{\cal P}_{ij}\right].
$$

\par ${\mathcal J}^{(4)}$: $(\mu,\eps)=(4,5)$
\beqan
&~&4\pi\eps_0 c^3{\mathcal J}^{(4)}=\left\{ \frac{2}{3c^2} (\ddot{\msf}+ \frac{1}{6c^2}\ddot{\ddot{\latens}}-\frac{1}{18c^2}\ddot{\ddot{\Nsf}}^{(3,2)})_i \right.\nonumber\\
&+&\left. \frac{1}{20c^4}(\tdot{{\cal M}}+\frac{1}{4c^2}
 \ddot{\tdot{\widetilde{\latens}}} -\frac{1}{24c^2}
 \ddot{\tdot{\widetilde{\Nsf}}}^{(4,2)})_{ij}(\tdot{{\cal M}}+\frac{1}{4c^2}
  \ddot{\tdot{\widetilde{\latens}}} -\frac{1}{24c^2}\ddot{\tdot{\widetilde{\Nsf}}}^{(4,2)})_{ij}\right.\nonumber\\
&+&\left. \frac{2}{945c^6}\ddot{\ddot{{\cal M}}}_{ijk}\ddot{\ddot{{\cal M}}}_{ijk}+\frac{1}{18144 c^8}\ddot{\tdot{\cal M}}_{ijkl}\ddot{\tdot{\cal M}}_{ijkl} \right.\nonumber\\
&+&\left. \frac{2}{3}(\ddot{\psf}-\frac{1}{c^2}\tdot{\Tsf}-\frac{1}{32c^4}\ddot{\tdot{\latens}}[\Nsf_{sym}^{(4,1)}]+\frac{1}{20c^4}\tdot{\tdot{\latens}}[\Pi^{(3)}]+\frac{1}{96c^4}\ddot{\tdot{\Nsf}}^{(4,3)})_i \right.\nonumber\\
&~&\left.(\ddot{\psf}-\frac{1}{c^2}\tdot{\Tsf}-\frac{1}{32c^4}\ddot{\tdot{\latens}}[\Nsf_{sym}^{(4,1)}]+\frac{1}{20c^4}\tdot{\tdot{\latens}}[\Pi^{(3)}]+\frac{1}{96c^4}\ddot{\tdot{\Nsf}}^{(4,3)})_i \right.\nonumber\\
&+&\left.\frac{1}{20c^2}(\tdot{{\cal P}}-\frac{1}{c^2}\ddot{\ddot{\Tsf}})_{ij}(\tdot{{\cal P}}-\frac{1}{c^2}\ddot{\ddot{\Tsf}})_{ij}+\frac{2}{945c^4}(\ddot{\ddot{{\cal P}}}-\frac{1}{c^2}\ddot{\tdot{\Tsf}})_{ijk} (\ddot{\ddot{{\cal P}}}-\frac{1}{c^2}\ddot{\tdot{\Tsf}})_{ijk} \right.\nonumber\\
&+&\left. \frac{1}{2^5 \times 3^4 \times 7}\ddot{\tdot{{\cal P}}}_{ijkl}\ddot{\tdot{{\cal P}}}_{ijkl}+\frac{1}{4 \times 3^3 \times 5^3 \times 77}\tdot{\tdot{{\cal P}}}_{ijklq}\tdot{\tdot{{\cal P}}}_{ijklq} \right\}_4
\eeqan
Only the fourth order terms must be kept from this expansion. This could be done by eliminating the useless terms, but, for obvious reasons, a detailed description is prefered here. The table with reduced moments is examined and the terms corresponding to the considered approximation are retained from the expressions representing the squares of these tensors. The result is:
\beqan
&~&4\pi\eps_0 c^3{\mathcal J}^{(4)}
=\frac{2}{3} \left(\ddot{\vec{p}}-\frac{1}{c^2}\tdot{\vec{\Tsf}} \right)^2\nonumber\\
&+& \frac{4}{3} \ddot{p}_i \left( - \frac{1}{32c^4}\ddot{\tdot{\latens}}[\Nsf_{sym}^{(4,1)}]+\frac{1}{20c^4}\tdot{\tdot{\latens}}[\Pi^{(3)}]+\frac{1}{96c^4}\ddot{\tdot{\Nsf}}^{(4,3)}\right)_i \nonumber\\
&~&+ \frac{1}{20c^2}\tdot{{\cal P}}_{ij}\tdot{{\cal P}}_{ij} - \frac{1}{10c^4}\tdot{{\cal P}}_{ij}\ddot{\ddot{\Tsf}}_{ij}+ 
\frac{2}{945c^4}\ddot{\ddot{{\cal P}}}_{ijk}\ddot{\ddot{{\cal P}}}_{ijk} \nonumber\\
&~&+\frac{2}{3c^2}\ddot{\vec{m}}^2 + \frac{4}{3c^2}\ddot{m}_i
 \left( \frac{1}{6c^2} \ddot{\ddot{\latens}} - \frac{1}{18c^2} 
 \ddot{\ddot{\Nsf}}^{(3,2)} \right)_i + \frac{1}{20c^4}\tdot{{\cal M}}_{ij}
 \tdot{{\cal M}}_{ij}
\eeqan
Because
\beqan
\fl&~&\Nsf^{(4,1)}_{qqi}-\frac{4}{5}\dot{\Pi}_{qqi}=
\frac{5}{2}\Lambda_i[\Nsf^{(4,1)}_{\scriptstyle sym}]-4\dot{\Lambda}_i[\Pi^{(3)}]
-\frac{5}{6}\Nsf^{(4,3)}_i=
\frac{2}{7}\intl_{\mathcal D}\big[2\xi^2(\xivec\cdot\jvec)\xi_i-3\xi^4j_i\big]\rmd^3\xi,
\eeqan
and
\beqan
&~&\Lambda_i-\frac{1}{3}\Nsf^{(3,2)}_i=\frac{2}{5}\msf_{qqi},
\eeqan
the last expression of ${\mathcal J}^{(4)}$ is the same as the expression given in equation 
\eref{IBB}.

\section{ Conclusions}

\par The multipole expansions in Cartesian coordinates are not largely treated in 
literature or, if they are, many inaccuracies and uncertainties appear. In papers like 
 \cite{vr} and  \cite{Di} 
the expansion in Cartesian coordinates of the electromagnetic field is presented and 
applied to radiation. Their results were presented here as well, in a more systematic 
and concise way. They were compared to other results from literature, usually obtained 
by expansion in spherical coordinates. It was underlined the fact that the problem is 
not accurately treated everywhere, this fact leading to some confusions.
\par The main difficulty in the case of Cartesian coordinates is the procedure of 
reduction of the {\it n}th-order multipole tensors to symmetric traceless ones to obtain 
in this way a description of multipoles in terms of irreducible rotation group representations. 
This procedure was presented here, in Chapter 3. The transformations implied in the 
reduction procedure were defined such that the electromagnetic potentials are altered 
only by gauge transformations. This implied a specific feature of the dynamic case: 
the redefinitions of the multipole tensors in the lower $n<N$ orders, induced by 
the reduction of tensors in a given order $N$.

\par The following problems are left for a future discussion and they 
could be some possible research topics:
\begin{itemize}
\item completely systemizing all types of moments grouped, in the different orders, 
in reduced moments;
\item defining the singular distributions associated to toroid moments and 
their physical meaning;
\item study of the implication of the existence of interactions associated
 to toroid moments and the setting, according to correct criteria, of 
the different terms contributions; 
\item making  symbolic programs which realize a reduction scheme of the type presented in 
Appendix for the general case.
\end{itemize}

\appendix

\section{Equivalent multipole expansions}
Following the basic ideas from \cite{vr}, we present in this Appendix a 
somehow different procedure for justifying the reduction scheme presented in Section 7.

Let the Taylor expansion of the $\delta-$ function:
\beqan
\fl&~&
\delta(\rvec-\rvec')=\suml_{n\geq 0}\frac{(-1)^n}{n!}x^{'}_{i_1}\dots x^{'}_{i_n}\d_{i_1}
\dots x^{'}_{i_n}\delta(\rvec).\eeqan
Applied to the current and charge distributions, it gives the following results:
\beqan
\fl&~&\jvec(\rvec,t)=\int \jvec(\rvec',t)\delta(\rvec-\rvec')\rmd^3x'=\suml_{n\geq 0}\frac{(-1)^n}{n!}
\int x'_{i_1}\dots x'_{i_n}\jvec(\rvec',t)\rmd^3x'\d_{i_1}\dots \d_{i_n}
\delta(\rvec),\eeqan
\beqan
\fl&~&\rho(\rvec,t)=\suml_{n\geq 0}\frac{(-1)^n}{n!}\int x'_{i_1}\dots x'_{i_n}\rho(\rvec',t)
\rmd^3x'\d_{i_1}\dots \d_{i_n}\delta(\rvec).
\eeqan
The electric multipole moments are introduced by
$$\psf_{i_1\dots i_n}=\int  x'_{i_1}\dots x'_{i_n}\rho(\rvec',t)\rmd^3x'$$
such that:
\beqan
\fl&~&\rho(\rvec,t)=\suml_{n\geq 0}\frac{(-1)^n}{n!}\psf_{i_1\dots i_n}(t)\d_{i_1}\dots \d_{i_n}
\delta(\rvec)=\suml_{n\geq 0}\frac{(-1)^n}{n!}\psf^{(n)}\vert\vert\nablav^n
\delta(\rvec).
\eeqan
Considering the continuity equation for electric charge, we may write the equation
$$
j_i=\nablav'(x'_i\jvec)+x'_i\frac{\d\rho}{\d t}.
$$
Let 
\beqan
\fl    &~&a^{(n)}_i=\int x'_{i_1}\dots x'_{i_n}j_i\d_{i_1}\dots \d_{i_n}\delta(\rvec)\\
  \fl  &~&= \int x'_{i_1}\dots x'_{i_n}\nablav'(x'_i\jvec)\rmd^3x'\d_{i_1}\dots \d_{i_n}\delta(\rvec)
+\left[\dot{\psf}^{(n+1)}(t)\vert\vert\nablav^n\right]_i\delta(\rvec)\\
   \fl&~&=-\int x'_i\jvec\cdot\nablav'(x'_{i_1}\dots x'_{i_n})\d_{i_1}\dots \d_{i_n}\delta(\rvec)+
\left[\dot{\psf}^{(n+1)}(t)\vert\vert\nablav^n\right]_i\delta(\rvec)\\
   \fl&~&=-n\int x'_{i_1}\dots x'_{i_{n-1}}x'_ij_{i_n}\rmd^3x'\d_{i_1}\dots \d_{i_n}\delta(\rvec)+
\left[\dot{\psf}^{(n+1)}(t)\vert\vert\nablav^n\right]_i\delta(\rvec)\\
   \fl&~&=-n\int x'_{i_1}\dots x'_{i_{n-1}}\left(x'_ij_{i_n}-x'_{i_n}j_i\right)\rmd^3x'
\d_{i_1}\dots \d_{i_n}\delta(\rvec)-a^{(n)}_i
+\left[\dot{\psf}^{(n+1)}(t)\vert\vert
\nablav^n\right]_i\delta(\rvec).
\eeqan
One obtains:
\beqan
\fl&~&a^{(n)}_i=-\frac{n}{n+1}\eps_{ii_nk}\int x'_{i_1}\dots x'_{i_{n-1}}\left(\rvec'\times
\jvec\right)_k\rmd^3x'\d_{i_1}\dots \d_{i_n}\delta(\rvec)\\
\fl&~&+\frac{1}{n+1}
\left[\dot{\psf}^{(n+1)}(t)\vert\vert\nablav^n\right]_i\delta(\rvec).
\eeqan
The magnetic multipole tensor is introduced by:
\beqan
\fl&~&\msf_{i_1\dots i_n}=\frac{n}{n+1}\int x'_{i_1}\dots x'_{i_{n-1}}
\left(\rvec'\times\jvec\right)_{i_n}\rmd^3x',\eeqan
and one may write
\beqan
\fl&~&a^{(n)}_i=-\eps_{ilk}\d_l\msf_{i_1\dots i_{n-1}k}
 \d_{i_1}\dots \d_{i_{n-1}}\delta(\rvec)+\frac{1}{n+1}
\left[\dot{\psf}^{(n+1)}(t)\vert\vert\nablav^n\right]_i\delta(\rvec)\eeqan
with the result
\beqan
\fl&~&    j_i(\rvec,t)=\suml_{n\geq 1}\frac{(-1)^{n-1}}{n!}\eps_{ilk}\d_l
\left[\msf^{(n)}\vert\vert\nablav^{n-1}\right]_k\delta(\rvec)\\
\fl&~&+
\suml_{n\geq 1}\frac{(-1)^{n-1}}{n!}\left[\dot{\psf}^{(n)}(t)
\vert\vert\nablav^{n-1}\right]_i\delta(\rvec).
\eeqan
Some invariance properties of the field  are established in the following. 
\par $\bullet$ By introducing in the expression of 
$\jvec$ the tensor $\msf^{(n)}_{(\scriptstyle L)}$ defined by the 
equation \eref{subst1}, and  using the relation 
$\eps_{ilk}\eps_{i_{\la}kq}=-\delta_{ii_{\la}}\delta_{lq}+\delta_{iq}
\delta_{i_{\la}q}$,
one obtains
\beqan
\fl&~&j_i=j_i\left( \msf^{(n)}\rightarrow\msf^{(n)}_{\scriptstyle (L)}\right)+
\frac{(-1)^{n-1}}{n!n}\suml^{n-1}_{\la=1}\eps_{ilk}\eps_{i_{\la}kq}{\mathcal N}^{(\la)}
_{i_1\dots i_{n-1}q}\big[\Lsf^{(n)}\big]\d_l\d_{i_1}\dots\d_{i_{n-1}}\delta(\rvec)\\
\fl   &~&=j_i\left(\msf^{(n)}\rightarrow \msf^{(n)}_{\scriptstyle (L)}\right)+
\frac{(-1)^n(n-1)}{n!n}\left[{\mathcal N}\big[\Lsf^{(n)}\big]\vert\vert\nablav^{n-2}
\right]_l\d_l\d_i\delta(\rvec)\\
   \fl&~&-\frac{(-1)^n(n-1)}{n!n}\left[{\mathcal N}\big[\Lsf^{(n)}\big]
   \vert\vert\nablav^{n-2}\right]_i\Delta\delta(\rvec).
\eeqan
Considering the corresponding  results for the potentials,
the presence of a term like $f(t)\Delta\delta(\rvec)$ in the expansion of  $\jvec$
 leads to a term in  $\Avec$ or in $\Phi$  which contains the expression
$$\Delta\frac{f(t-R/c)}{r}=\frac{1}{c^2}\frac{\ddot{f}(t))}{r},\;\; r\neq 0$$ 
Therefore, regarding the contribution to the vector potential, the $\jvec$ expansion 
is equivalent to:
\beqan
  \fl &~&j^{\,'}_i=j_i\left(\msf^{(n)}\rightarrow \msf^{(n)}_{\scriptstyle (L)}\right)
  -\frac{(-1)^n(n-1)}{n!nc^2}
\left[\ddot{\mathcal N}\big[\Lsf^{(n)}\big]\vert\vert\nablav^{n-2}\right]_i\delta(\rvec)\\
   \fl&~&+\frac{(-1)^n(n-1)}{n!n}\left[{\mathcal N}\big[\Lsf^{(n)}\big]\vert\vert\nablav^{n-2}\right]_l
\d_l\d_i\delta(\rvec).\eeqan
The last term in the precedent expression produces in  $\Avec$ an additional  term like
$\nablav\Psi(\rvec,t)$ which, as one sees bellow, is a  contribution from a gauge 
transformation of the potentials.
\par Let  the transformation 
$$\psf_{i_1\dots i_{n-1}}\longrightarrow\psf'_{i_1\dots i_{n-1}}=
\psf_{i_1\dots i_{n-1}}-\frac{n-1}{c^2n^2}
\dot{\mathcal N}_{i_1\dots i_{n-1}}\big[\Lsf^{(n)}\big].$$
By substituting   $\psf_{i_1\dots i_{n-1}}$ in the $\jvec$ and  $\rho$ expansions by 
\beq\label{tr-P(n-1)}
\psf_{i_1\dots i_{n-1}}=\psf'_{i_1\dots i_{n-1}}+\frac{n-1}{c^2n^2}
\dot{\Nsf}_{i_1\dots i_{n-1}}.
\eeq
one gets:
\beqan
&~& \fl j^{'}_i= 
j_i\left(\begin{array}{ccc}\scriptsize{ \msf^{(n)}\rightarrow \msf^{(n)}_{\scriptstyle 
(L)}}\\
 \scriptsize{\psf^{(n-1)}\rightarrow \psf'^{(n-1)}}\end{array}\right)
+\frac{(-1)^n(n-1)}{n!n}\left[{\mathcal N}\big[\Lsf^{(n)}\big]\vert\vert\nablav^{n-1}
\right]_l\d_l\d_i\delta(\rvec)
\eeqan
\beqan
\fl   &~&\rho=
\rho\left({\scriptsize\psf^{(n-1)}}\rightarrow \scriptsize{\psf'^{(n-1)}}\right)
-\frac{(-1)^{n}(n-1)}{n!nc^2}\left[\dot{\mathcal N}\big[\Lsf^{(n)}\big]
\vert\vert\nablav^{n-1}\right]\delta(\rvec).
\eeqan

It is easy to see that the  potentials $(\widetilde{\Phi},\;\widetilde{\Avec})$ 
corresponding to  the new densities $(\tilde{\rho},\;\widetilde{\jvec})$ with 
$$
\tilde{\rho}=\rho\left({\scriptsize\psf^{(n-1)}}\rightarrow 
\scriptsize{\psf'^{(n-1)}}\right),\;
\widetilde{\jvec}=\jvec\left(\begin{array}{ccc}\scriptsize{ \msf^{(n)}
\rightarrow \msf^{(n)}_{\scriptstyle(L)}}\\
 \scriptsize{\psf^{(n-1)}\rightarrow \psf'^{(n-1)}}\end{array}\right)
$$
differ from the original one by a gauge transformation.
\par We point out that the physical equivalence of the two distributions $(\rho,\;
\jvec)$ and $(\widetilde{\rho},\;\widetilde{\jvec})$ is true only for the field 
in the exterior of the domain ${\mathcal D}$ including the support of the 
charges and currents. 
\par $\bullet$ Let 
$$
\psf_{{\scriptstyle (L)}i_1\dots i_n}=\psf_{i_1\dots i_n}-
\frac{1}{n}\suml^{n-1}_{\la=1}\eps_{i_{\la}i_nq}{\mathcal N}^{(\la)}_{i_1\dots i_{n-1}q}
\big[\Lsf^{(n)}\big].
$$
It is a simple matter to see that the transformation $\psf^{(n)}\longrightarrow 
\psf^{(n)}_{\scriptstyle N}$ does not alter the density $\rho$ but
\beqan
\fl&~& j_i=j_i\big(\psf^{(n)}\longrightarrow \psf^{(n)}_{\scriptstyle (L)}\big)
-\frac{(-1)^{n-1}(n-1)}{n!n}\eps_{liq}\d_l\left[\dot{\mathcal N}_{i_1\dots i_{n-2}}
\big[\Lsf\big]\,\vert\vert\nablav^{n-2}\right]_q\delta(\rvec).
\eeqan 
\par The difference between $j_i[\psf^{(n)}]$ and $j_i[\psf^{(n)}\rightarrow 
\psf^{(n)}_{\scriptstyle (L)}]$ may be eliminated by the transformation
$$
 \msf_{i_1\dots i_{n-1}}\longrightarrow \msf'_{1_1\dots i_{n-1}}\msf_{i_1\dots i_{n-1}}
 +\frac{n-1}{n^2}\dot{\mathcal N}_{i_1\dots i_{n-1}}[\Lsf]
$$
 and so
 $$
 j_i=j_i\left(\begin{array}{ccc}{\scriptsize \psf^{(n)}\rightarrow 
 \psf^{(n)}_{\scriptstyle L}}\\
 {\scriptsize \msf^{(n-1)}\rightarrow \msf'^{(n-1)}}\end{array}\right).
 $$
\par $\bullet$ Let  the transformation \eref{subst3} of the magnetic tensor
$$\msf_{i_1\dots i_n}\longrightarrow\msf_{{\scriptstyle (S)}i_1\dots i_n}=
\msf_{i_1\dots i_n}-\suml_{D(i)}\delta_{i_1i_2}
\Lambda_{i_3\dots i_n}\big[\Ssf^{(n)}\big]$$
where $\Ssf^{(n)}$ is a fully symmetric tensor and the operator $\Lambda$ is 
used as in equation \eref{defS}  for detracing an arbytrary  fully symmetric tensor.
\par Introducing the tensor $\msf^{(n)}_{\scriptstyle (S)}$ in the expansion of $\jvec$ we may write
$$j_i=j_i\big({\scriptsize \msf^{(n)}\rightarrow{\msf}^{(n)}_{\scriptstyle (S)}}\big)
+\frac{(-1)^{n-1}}{n!}\eps_{ilk}\d_l\suml_{D(i)}\delta_{i_1i_2}\Lambda_{i_3\dots 
i_{n-1}k}\big[\Ssf^{(n)}\big]\d_{i_1}\dots \d_{i_{n-1}}\delta(\rvec)$$ 
$\Lambda$ being a fully symmetric tensor. 
In the sum there are  $n-1$ terms with $\delta_{ki_j},\; j=1\dots n-1$ but 
$\delta_{ki_j}\d_{i_j}=\d_k$ and $\eps_{ilk}\d_l\d_k=0$ such that  these terms  
give a null result. In the remaining terms  we have the factors like 
 $\delta_{i_ji_q},\;j,q=1\dots n-1$ 
resulting $\delta_{i_ji_q}\d_{i_j}\d_{i_q}=\Delta$.
The number of these terms is   $C^2_{n-1}=(n-1)(n-2)/2$, such that
\beqan
  \fl  &~&j_i=j_i\big({\scriptsize \msf^{(n)}\rightarrow \msf}^{(n)}_{\scriptstyle (S)}
  \big)
  +\frac{(-1)^{n-1}(n-1)(n-2)}
{2n!}\eps_{ilk}\d_l\Lambda_{i_1\dots i_{n-3}k}\big[\Ssf^{(n)}\big]
\d_{i_1}\dots\d_{i_{n-3}}\Delta\delta(\rvec)\\
  \fl &~&=j_i\big({\scriptsize \msf^{(n)}\rightarrow \msf^{(n)}_{\scriptstyle (S)}}\big)+
\frac{(-1)^{n-1}(n-1)(n-2)}{2n!}\left\{\nablav\times\left[\Lambda\big[\Ssf^{(n)}\big]
\vert\vert\nablav^{n-3}\right]\right\}_i\Delta\delta(\rvec).
\eeqan
This relation allows the introduction of an equivalent  current density regarding the 
vector potential:
\beqan
  \fl &~&j_i"=j_i\big({\scriptsize \msf^{(n)}\rightarrow \msf^{(n)}_{\scriptstyle (S)}}\big)+
\frac{(-1)^{n-1}(n-1)(n-2)}{2n!c^2}\left\{\nablav\times\left[\ddot{\Lambda}
\big[\Ssf^{(n)}\big]\vert\vert\nablav^{n-3}\right]\right\}_i\delta(\rvec).
\eeqan
The suplimentary term is eliminated by the transformation
$$
\msf^{(n-2)}\rightarrow\msf'^{(n-2)}=\msf^{(n-2)}+\frac{n-2}{2nc^2}\ddot{\Lambda}
\big[\Ssf^{(n)}\big]
$$
such that 
\beqan
\fl&~&\jvec"=\jvec\left(\begin{array}{ccc}
\scriptsize{ \msf^{(n)}\rightarrow \msf^{(n)}_{\scriptstyle (S)}}\\
\scriptsize{ \msf^{(n-2)}\rightarrow\msf'^{(n-2)}}\end{array}\right).
\eeqan
\par $\bullet$ Let 
\beqan
\fl&~&\psf_{{\scriptstyle (S)}i_1\dots i_n}=\psf_{i_1\dots i_n}-
\suml_{D(i)}\delta_{i_1i_2}\Lambda_{i_3\dots i_n}[\Ssf^{(n)}].
\eeqan
One gets
\beqan
 \fl&~&j_i=j_i\left(\psf^{(n)}\rightarrow \psf^{(n)}_{\scriptstyle S}\right)
+\frac{(-1)^{n-1}}{n!}\left[\suml_{D(i)}\delta_{i_1i_2}
\dot{\Lambda}_{i_3\dots i_{n-1}i}[\Ssf^{(n)}]\right]\d_{i_1\dots 
i_{n-1}}\delta(\rvec)\nonumber\\
   \fl&~&=j_i\left(\psf^{(n)}\rightarrow \psf^{(n)}_{\scriptstyle S}\right)
+\frac{(-1)^{n-1}(n-1)}{n!}\left[\dot{\Lambda}[\Ssf^{(n)}]\vert\vert\nablav^{n-2}\right]
\d_i\delta(\rvec)\nonumber\\
   \fl&~&+\frac{(-1)^{n-1}(n-1)(n-2)}{2n!}\left[\dot{\Lambda}[\Ssf^{(n)}]
\vert\vert\nablav^{n-3}\right]_i\Delta\delta(\rvec)'
\eeqan
So, we obtain an equivalent current density
\beqan
 \fl&~&j'_i=
   j_i\left(\psf^{(n)}\rightarrow \psf^{(n)}_{\scriptstyle S}\right)
+\frac{(-1)^{n-1}(n-1)}{n!}\left[\dot{\Lambda}[\Ssf^{(n)}]\vert\vert\nablav^{n-2}\right]
\d_i\delta(\rvec)\nonumber\\
   \fl&~&+\frac{(-1)^{n-1}(n-1)(n-2)}{2n!c^2}\left[\tdot{\Lambda}[\Ssf^{(n)}]
\vert\vert\nablav^{n-3}\right]_i\delta(\rvec).
\eeqan
For the charge density we have
\beqan
\fl&~&\rho=\rho\left(\psf^{(n)}\rightarrow \psf^{(n)}_{\scriptstyle S}\right)
+\frac{(-1)^n}{n!}\suml_{D(i)}\delta_{i_1i_2}\Lambda_{i_3\dots i_n}[\Ssf^{(n)}]
\d_{i_1}\dots \d_{i_n}\delta(\rvec).\eeqan
Because in the last sum there are $C^2_n$ terms like $\delta_{i_ji_q}$ with equal 
contributions, one obtains 
$$
\rho=\rho\left(\psf^{(n)}\rightarrow \psf^{(n)}_{\scriptstyle S}\right)
+\frac{(-1)^nn(n-1)}{2n!}\left[{\Lambda}[\Ssf^{(n)}]\vert\vert\nablav^{n-2}
\right]\Delta\delta(\rvec)$$
and we may introduce the equivalent density
$$
\rho'=\rho\left(\psf^{(n)}\rightarrow \psf^{(n)}_{\scriptstyle S}\right)
+\frac{(-1)^nn(n-1)}{2n!c^2}\left[\ddot{\Lambda}[\Ssf^{(n)}]\vert\vert\nablav^{n-2}
\right]\delta(\rvec)
$$
Let us define 
$$
\psf^{"(n-2)}=\psf^{(n-2)}+\frac{n-2}{2nc^2}\ddot{\Lambda}[\Ssf^{(n)}].
$$
It is easy to see that
$$j'_i=j_i\left(\begin{array}{ccc}{\scriptsize \psf^{(n)}\rightarrow \psf^{(n)}_
{\scriptstyle S}}\\
{\scriptsize \psf^{(n-2)}\rightarrow \psf^{"(n-2)}}\end{array}\right)
+\frac{(-1)^{n-1}(n-1)}{n!}\big[\dot{\Lambda}[\Ssf^{(n)}]\,\vert\vert
\nablav^{n-2}\big]\d_i\delta(\rvec)$$
and

$$\rho'=\rho\left(\begin{array}{ccc}{\scriptsize \psf^{(n)}\rightarrow \psf^{(n)}_
{\scriptstyle S}}\\
{\scriptsize \psf^{(n-2)}\rightarrow \psf^{"(n-2)}}\end{array}\right)
+\frac{(-1)^n(n-1)}{n!c^2}\big[\ddot{\Lambda}[\Ssf^{(n)}]\vert\vert\nablav^{n-2}
\big]\delta(\rvec).$$
So, the densities $(\rho',\jvec^{\,'})$ and 
$$\left\{\rho\left(\begin{array}{ccc}{\scriptsize \psf^{(n)}\rightarrow \psf^{(n)}_
{\scriptstyle S}}\\
{\scriptsize \psf^{(n-2)}\rightarrow \psf^{"(n-2)}}\end{array}\right),
\jvec\left(\begin{array}{ccc}{\scriptsize \psf^{(n)}\rightarrow \psf^{(n)}_
{\scriptstyle S}}\\
{\scriptsize \psf^{(n-2)}\rightarrow \psf^{"(n-2)}}\end{array}\right)
 \right\}$$
 are equivalent because the corresponding potentials differ only by a gauge transformation.

\section{The Reduction Scheme}
 
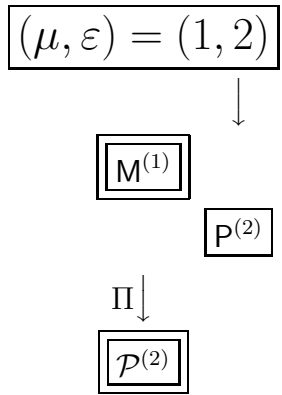
\beqan
\begin{CD}
\fl\framebox{\Large{$(\mu,\eps)=(1,2)$}}\\
@VVV\\
\fl\framebox{\framebox{$\msf^{(1)}$}}\\
\framebox{$\psf^{(2)}$}\\
\fl@V\hbox{$\Pi$}VV\\
\fl\framebox{\framebox{$\mathcal{P}^{(2)}$}}\\
\fl\mbox{$\mbox{Note.\,\,}\mathcal{P}_{ij}=\psf_{ij}-\frac{1}{3}\psf_{qq}\delta_{ij}$,}\\
\fl\mbox{$\mbox{ no\; extra-gauge\; 
changes\; for}\; \Avec\; \mbox{and}\; \Phi \;\mbox{for}\; \psf^{(2)}
\rightarrow \mathcal{P}^{(2)}$}\\
\end{CD}
\eeqan

\par \,\,\,\,
\beqan
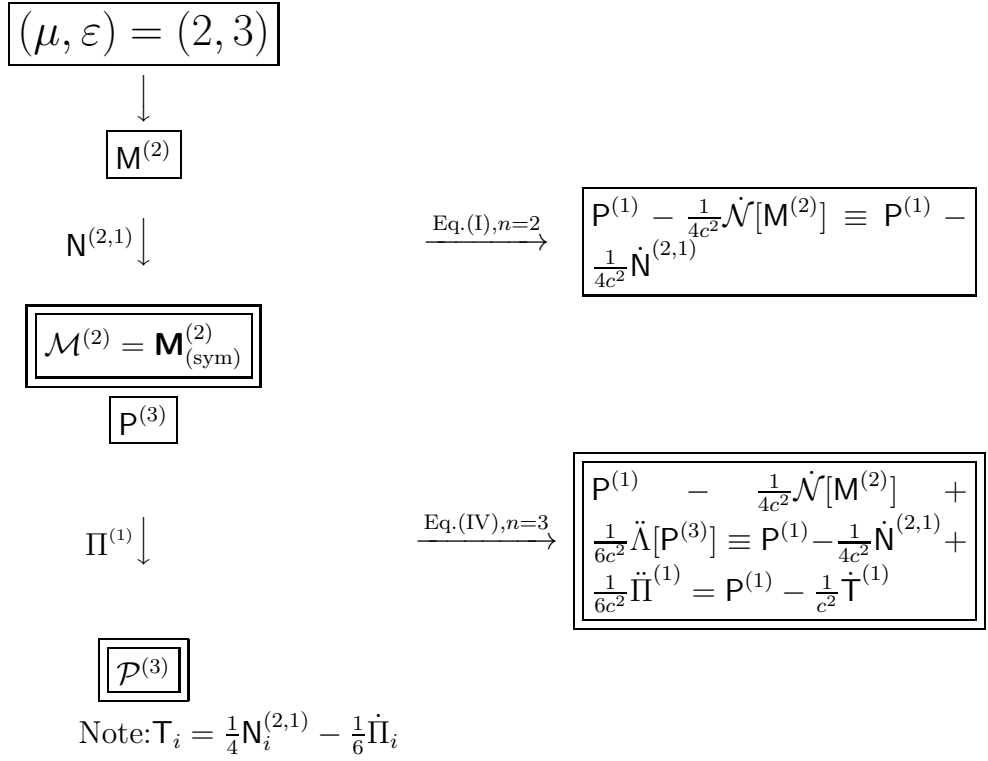
\begin{CD}
\fl\framebox{\Large{$(\mu,\eps)=(2,3)$}}\\
\fl@VVV\\
\fl\framebox{$\msf^{(2)}$}\\
 \fl@V \hbox{$\Nsf^{(2,1)}$}VV @>\mbox{\scriptsize Eq.(I)},
n=2>>
\framebox{\parbox{5.0cm}{$ 
\psf^{(1)}-\frac{1}{4 c^2}\dot{\mathcal{N}}[\msf^{(2)}]\equiv
\psf^{(1)}-\frac{1}{4 c^2}\dot{\Nsf}^{(2,1)}$}}\\
\fl\framebox{\framebox{${\mathcal M}^{(2)}=\msfsim^{(2)}$}}\\
\fl\framebox{$\psf^{(3)}$}\\
 \fl@V\hbox{$\Pi$}^{(1)}VV @>\mbox{\scriptsize Eq.(IV)},
n=3>>
\framebox{\framebox{\parbox{5.0cm}{$\psf^{(1)}-\frac{1}{4c^2}\dot{\mathcal N}[\msf^{(2)}]
+\frac{1}{6c^2}\ddot{\Lambda}[\psf^{(3)}]
\equiv
\psf^{(1)}-\frac{1}{4 c^2}\dot{\Nsf}^{(2,1)}+\frac{1}{6c^2}\ddot
{\hbox{$\Pi$}}^{(1)}=\psf^{(1)}-\frac{1}{c^2}\dot{\Tsf}^{(1)}$}}}\\
\fl\framebox{\framebox{$\mathcal{P}^{(3)}$}}\\
\mbox{$\mbox{Note:}\Tsf_i=\frac{1}{4}\Nsf^{(2,1)}_i-\frac{1}{6}\dot{\Pi}_i$}\\
\end{CD}
\eeqan
\vspace{-5.0cm}
\beqan
\begin{CD}
\fl\framebox{\Large{$(\mu,\eps)=(3,4)$}}\\
 @VVV\\
\fl\framebox{$\msf^{(3)}$}\\
 \fl@V
\hbox{$\Nsf^{(3,1)}$}VV @>\mbox{\scriptsize Eq.(I)},
n=3>>
\framebox{\parbox{4.0cm}{$ \mathcal{P}^{(2)}-\frac{2}{9 c^2}
\dot{\mathcal N}[\msf^{(3)}]\equiv
\mathcal{P}^{(2)}-\frac{2}{9 c^2}\dot{\Nsf}^{(3,1)}$}}\\
\fl\framebox{$\msfsim^{(3)}$}\\
\fl @V\hbox{$\Lambda$}^{(1)}VV @>Eq.\mbox{\scriptsize}(III),n=3>>
\framebox{\parbox{4.0cm}{$ \msf^{(1)}+\frac{1}{6 c^2}\ddot{\Lambda}
[\msf^{(3)}_{sym}]\equiv
\msf^{(1)}+\frac{1}{6 c^2}\ddot{\Lambda}^{(1)}$}}\\
\fl\framebox{\framebox{$\mathcal{M}^{(3)}$}}\\
\fl\framebox{$\psf^{(4)}$}\\
 \fl@V\hbox{$\Pi$}^{(2)}VV @>\mbox{\scriptsize Eq.(IV)},
n=4>>
\framebox{\parbox{4.5cm}{$
\mathcal{P}^{(2)}-\frac{2}{9 c^2}\dot{\mathcal N}[\msf^{(3)}]+\frac{1}{4c^2}\ddot{\Lambda}[
\psf^{(4)}]\equiv
\mathcal{P}^{(2)}-\frac{2}{9 c^2}\dot{\Nsf}^{(3,1)}+\frac{1}{4c^2}\ddot{\hbox
{$\Pi$}}^{(2)}$}}\\
\fl\framebox{\framebox{$\mathcal{P}^{(4)}$}}\\
\fl\framebox{\parbox{4.0cm}{$\mathcal{P}^{(2)}-\frac{2}{9 c^2}\dot{\mathcal N}[\msf^{(3)}]+
\frac{1}{4c^2}\ddot{\Lambda}[\psf^{(4)}]$}}\\
 \fl@V\hbox{$ \dot{\Nsf}^{(3,2)}$}
 VV @>\mbox{\scriptsize Eq.(II)},n=3>>
\framebox{
\parbox{4.5cm}{$ \msf^{(1)}+\frac{1}{6c^2}\ddot{\Lambda}[\msf^{(3)}_{\scriptstyle sym}]
-\frac{1}{18c^2}\ddot{\mathcal{N}}^2[\msf^{(3)}]\equiv
\framebox{\parbox{3.5cm}{$\msf^{(1)}+\frac{1}{6c^2}\ddot{\Lambda}^{(1)}
-\frac{1}{18c^2}\ddot{N}^{(3,2)}$}}
$}}\\
\fl\framebox{$\mathcal{P}^{(2)}-\frac{2}{9 c^2}\dot{\mathcal N}_{\scriptstyle 
\rm{sym}}[\msf^{(3)}]+\frac{1}{4c^2}\ddot{\Lambda}[\Pi^{(4)}$}\\
\fl @V\hbox{$\ddot{\Lambda}[\Pi^{(2)}]$}VV \\
\framebox{\framebox{\parbox{5.0cm}{$\mathcal{P}^{(2)}-\frac{2}{9 c^2}\dot{\widetilde
{\mathcal N}}[\msf^{(3)}]
+\frac{1}{4c^2}\ddot{\widetilde{\Lambda}}[\psf^{(4)}]={\mathcal P}^{(2)}
-\frac{1}{c^2}\dot{\Tsf}^{(2)}$}}}\\
\fl\mbox{$\mbox{Note.}\;\; \Nsf^{(3,1)}_{sym}=\widetilde{\Nsf}^{(3,1)}=\mbox{traceless}$}\\
\mbox{$\Tsf_{ij}=\frac{2}{9}\widetilde{\Nsf}^{(3,1)_{ij}}-\frac{1}{4}\dot{\widetilde{\Pi}}_{ij}$}\\
\fl\mbox{$\mbox{Transf.}\; \Pi^{(2)}\rightarrow \widetilde{\Pi^{(2)}}$}\\
\fl\mbox{$ \mbox{produces}\;$}
\mbox{$\mbox{only\; a}$}\\
\fl\mbox{$\mbox{ gauge\;transformation}$}\\
\end{CD}
\eeqan
\framebox{$(\mu,\eps)=(4,5)$}
\beqan
\begin{CD}
\fl\framebox{$\msf^{(4)}$}\\
\fl @V\hbox{$\Nsf^{(4,1)}$}VV @>\mbox{\scriptsize Eq.(I)},
n=4>>
\framebox{\parbox{6.0cm}{$
\mathcal{P}^{(3)}-\frac{3}{16 c^2}\dot{\mathcal N}[\msf^{(4)}]\equiv
\mathcal{P}^{(3)}-\frac{3}{16 c^2}\dot{\Nsf}^{(4,1)}$}}\\
\fl\framebox{$\msfsim^{(4)}$}\\
 @V\hbox{$\Lambda$}^{(2)}VV @>Eq.(III),n=4>>
\framebox{\parbox{6.0cm}{$ 
{\mathcal M}^{(2)}+\frac{1}{4 c^2}\ddot{\Lambda}[\msf^{(4)}_{sym}]\equiv
{\mathcal M}^{(2)}+\frac{1}{4 c^2}\ddot{\Lambda}^{(2)}$}}\\
\fl\framebox{\framebox{$\mathcal{M}^{(4)}$}}\\
\fl\framebox{$\psf^{(5)}$}\\
 @V\hbox{$\Pi$}^{(3)}VV @>\mbox{\scriptsize Eq.(IV)},
n=5>>
\framebox{\parbox{6.0cm}{$ 
\mathcal{P}^{(3)}-\frac{3}{16 c^2}\dot{\mathcal N}[\msf^{(4)}]+
\frac{3}{10c^2}\ddot{\Lambda}[\psf^{(5)}]\equiv
\mathcal{P}^{(3)}-\frac{3}{16 c^2}\dot{\Nsf}^{(4,1)}+\frac{3}{10c^2}\ddot{\Pi}^{(3)}$}}\\
\fl\framebox{\framebox{$\mathcal{P}^{(5)}$}}\\
\fl\framebox{\parbox{6.0cm}{$\mathcal{P}^{(3)}-\frac{3}{16 c^2}\dot{\Nsf}^{(4,1)}+
\frac{3}{10c^2}\ddot{\Pi}^{(3)}$}}\\
 \fl @V\hbox{$ \dot{\Nsf}^{(4,2)}$}VV @>\mbox{\scriptsize Eq.(II)},
n=4>>
\framebox{\parbox{6.0cm}{$ 
\mathcal{M}^{(2)}+\frac{1}{4c^2}\ddot{\Lambda}[\msf^{(4)}_{\scriptstyle sym}]
-\frac{1}{24c^2}\ddot{\mathcal N}^2[\msf^{(4)}_{\scriptstyle sym}]\equiv
\mathcal{M}^{(2)}+\frac{1}{4c^2}\ddot{\Lambda}^{(2)}
-\frac{1}{24c^2}\ddot{\Nsf}^{(4,2)}$}}\\
\fl\framebox{$\mathcal{P}^{(3)}-\frac{3}{16 c^2}\dot{\Nsf}^{(4,1)}_{\scriptstyle
\rm{sym}}+\frac{3}{10c^2}\ddot{\Pi}^{(3)}$}\\
\fl @V\hbox{$\dot{\Lambda}[\Nsf^{(4,1)}_{\scriptstyle sym}],\;
  \ddot{\Lambda}[\Pi^{(3)}]$}VV @>\mbox{\scriptsize 
 Eq.(IV)},
n=3>>
\framebox{\parbox{6.0cm}{$\psf^{(1)}-\frac{1}{c^2}\dot{\Tsf}^{(1)}
-\frac{1}{32c^4}\dot{\ddot{\Lambda}}
[{\mathcal N}_{\scriptstyle sym}[\msf^{(4)}]]+\frac{1}{20c^4}\ddot{\ddot{\Lambda}}[
\psf^{(5)}]\equiv \psf^{(1)}-\frac{1}{c^2}\dot{\Tsf}^{(1)}
-\frac{1}{32c^4}\dot{\ddot{\Lambda}}[\Nsf^{(4,1)}_{\scriptstyle sym}]
+\frac{1}{20c^4}\ddot{\ddot{\Lambda}}[\Pi^{(3)}]
$}}\\
\fl\framebox{\framebox{\parbox{6.0cm}{$\mathcal{P}^{(3)}-\frac{3}{16 c^2}\dot{\widetilde{\Nsf}}^
{(4,1)}+\frac{3}{10c^2}\ddot{\widetilde{\Pi}}^{(3)}={\mathcal P}^{(3)}-
\frac{1}{c^2}\dot{\Tsf}^{(3)}$}}}\\
\fl\framebox{$\mathcal{M}^{(2)}+\frac{1}{4c^2}\ddot{\Lambda}{(2)}
-\frac{1}{24c^2}\ddot{\Nsf}^{(4,2)}$}\\
 \fl@V\hbox{$\ddot{\Nsf}^{(4,3)}$}VV @>
\mbox{\scriptsize Eq.(I)},n=2>>
\framebox{
\framebox{\parbox{6.0cm}{$
\psf^{(1)}-\frac{1}{c^2}\dot{\Tsf}^{(1)}
-\frac{1}{32c^4}\dot{\ddot{\Lambda}}[{\mathcal N}_{\scriptstyle sym}[\msf^{(4)}
]]+\frac{1}{20c^4}\ddot{\ddot{\Lambda}}[\Lambda[\psf^{(5)}]]
+\frac{1}{96c^4}\dot{\ddot{\mathcal N}}^3[\msf^{(4)}]\equiv
\psf^{(1)}-\frac{1}{c^2}\dot{\Tsf}^{(1)}
-\frac{1}{32c^4}\dot{\ddot{\Lambda}}[\Nsf^{(4,1)}_{sym}]
+\frac{1}{20c^4}\ddot{\ddot{\Lambda}}[\Pi^{(3)}]
+
\frac{1}{96c^4}\dot{\ddot{\Nsf}}^{(4,3)}$}}}\\
\fl\framebox{\framebox{$ \mathcal{M}^{(2)}+\frac{1}{4c^2}
\ddot{\Lambda}^{(2)}-\frac{1}{24c^2}\ddot{{\Nsf}}^{(4,2)}_{\scriptstyle sym} $}}\\
 \fl@V\hbox{$\Lambda...$}VV @>\mbox{\scriptsize no effect}>>
\framebox{}
\\
\fl\framebox{\framebox{$ \mathcal{M}^{(2)}+\frac{1}{4c^2}
\ddot{\widetilde{\Lambda}}^{(2)}-\frac{1}{24c^2}\ddot{\widetilde{\Nsf}}^{(4,2)}_{\scriptstyle sym} $}}\\
\end{CD}
\eeqan

\section{Results}

\begin{center}
Table I
\end{center}
\begin{tabular}{|r|l|l|l|l|}
\hline
&$(\mu,\eps)=(1,2)$ & $(\mu,\eps)=(2,3)$ & $(\mu,\eps)=(3,4)$ \\
\hline
$\widetilde{\psf}^{(1)}$ & $\psf^{(1)}:(\vec{p})$&$\psf^{(1)}-\frac{1}{c^2}\dot{\Tsf}^{(1)}$
 &$\psf^{(1)}-\frac{1}{c^2}\dot{\Tsf}^{(1)}$\\
\hline
$\widetilde{\psf}^{(2)}$ & ${\mathcal P}^{(2)}:\;{\mathcal P}_{ij}=\psf_{ij}-\frac{1}{3}
\psf_{qq}\delta_{ij}$&${\mathcal P}^{(2)}$ &${\mathcal P}^{(2)}-\frac{1}{c^2}\dot{\Tsf}^{(2)}$\\
\hline
$\widetilde{\psf}^{(3)}$ & &${\mathcal P}^{(3)}$ &${\mathcal P}^{(3)}$\\
\hline
$\widetilde{\psf}^{(4)}$ & $$& &${\mathcal P}^{(4)}$\\
\hline
$\widetilde{\psf}^{(5)}$ & $$& &\\
\hline
$\widetilde{\msf}^{(1)}$ & $\msf^{(1)}:(\vec{m})$&$\msf^{(1)}$ &$\msf^{(1)}+\frac{1}{6c^2}\ddot{\Lambda}^{(1)}-
\frac{1}{18c^2}\ddot{\Nsf}^{(3,2)}$\\
\hline
$\widetilde{\msf}^{(2)}$ &&${\mathcal M}^{(2)}$ &${\mathcal M}^{(2)}$\\
\hline
$\widetilde{\msf}^{(3)}$ & $$& &${\mathcal M}^{(3)}$\\
\hline
$\widetilde{\msf}^{(4)}$ & $$& &\\
\hline
\end{tabular}\\
\begin{center}
Table II
\end{center}
\begin{tabular}{|r|l|}
\hline
&$~~~~~~~~~~~~~~~~~~~~~~~~(\mu,\eps)=(4,5)$ \\
\hline
$\widetilde{\psf}^{(1)}$
 &$\psf^{(1)}-\frac{1}{c^2}\dot{\Tsf}^{(1)}-\frac{1}{32c^4}
\tdot{\Lambda}[\Nsf^{(4,1)}_{\scriptstyle sym}]+
\frac{1}{20c^4}\ddot{\ddot{\Lambda}}[\Pi^{(3)}]
+\frac{1}{96c^4}\tdot{\Nsf}^{(4,3)} $\\
\hline
$\widetilde{\psf}^{(2)}$ &${\mathcal P}^{(2)}-\frac{1}{c^2}\dot{\Tsf}^{(2)} $\\
\hline
$\widetilde{\psf}^{(3)}$ &${\mathcal P}^{(3)}-\frac{1}{c^2}\dot{\Tsf}^{(3)}$\\
\hline
$\widetilde{\psf}^{(4)}$ &${\mathcal P}^{(4)}$ \\
\hline
$\widetilde{\psf}^{(5)}$ &${\mathcal P}^{(5)}$ \\
\hline
$\widetilde{\msf}^{(1)}$ &$\msf^{(1)}+\frac{1}{6c^2}\ddot{\Lambda}^{(1)}-
\frac{1}{18c^2}\ddot{\Nsf}^{(3,2)}$ \\
\hline
$\widetilde{\msf}^{(2)}$ &${\mathcal M}^{(2)}+\frac{1}{4c^2}\ddot{\widetilde{\Lambda}}^{(2)}
-\frac{1}{24c^2}\ddot{\widetilde{\Nsf}}^{(4,2)}$\\
\hline
$\widetilde{\msf}^{(3)}$ &${\mathcal M}^{(3)}$\\
\hline
$\widetilde{\msf}^{(4)}$ &${\mathcal M}^{(4)}$\\
\hline
\end{tabular}
~~~~~~~~~~~~~~~~~~~~~~~~~~~~~~~~~~~~~~~~~~~~~~~~~~~~~~~~~~~~~~~~~~~~~~~~~~~~~~~~~~~~~~~~~~~~~~~~~~~~~~~~~~~~~~~~~~~~~~~~~~~~~~~~~~~~~~~~~~~~~~~~~~~~~~~~~~~~~~~~~~~~~~~~~~~~~~~~~~~~~~~~~~~~~~~~~~~~~~~~~~~~~~~~~~~~~~~~~~~~~~~~~~~~~~~~~~~~~~~~~~~~~~~~~~~~~~~~~~~~~~~~~~~
\vspace{2.0cm}

In the above formulas it was introduced the electric toroid moment of undefined order, as it results from the algorithm presented before:
$$
\Tsf^{(n)}=\frac{n}{(n+1)^2}\widetilde{\Nsf}^{(n+1,1)}-\frac{n}{2(n+2)}
\dot{\widetilde{\Pi}}^{(n)}
$$

\vspace{0.5cm}
\par

\end{document}